\documentclass[1p,authoryear]{elsarticle}

\usepackage{amsmath}
\usepackage{graphicx}
\usepackage{svg}
\usepackage{commath}  

\usepackage{color}
\usepackage[normalem]{ulem}
\usepackage{hyperref}
\usepackage{amsfonts}

\begin{document}

\begin{frontmatter}

\title{Crack-front model for adhesion of soft elastic spheres with chemical heterogeneity}

\begin{abstract}
Adhesion hysteresis can be caused by elastic instabilities that are triggered by surface roughness or chemical heterogeneity.
However, the role of these instabilities in adhesion hysteresis remains poorly understood because we lack theoretical and numerical models accounting for realistic roughness.
Our work focuses on the adhesion of soft elastic spheres with low roughness or weak heterogeneity,
where the indentation process can be described as a Griffith-like propagation of a nearly circular external crack.
We discuss how to describe the contact of spheres with chemical heterogeneity that leads to fluctuations in the local work of adhesion. We introduce a variational first-order crack-perturbation model and validate our approach using boundary-element simulations.
The crack-perturbation model faithfully predicts contact shapes and hysteretic force-penetration curves,
provided that the contact perimeter remains close to a circle and the contact area is simply connected.
Computationally, the crack-perturbation model is orders of magnitude more efficient than the corresponding boundary element formulation, allowing for realistic heterogeneity fields.
Furthermore, our crack-front formulation clarifies the connection of adhesion hysteresis to classic theories on pinning of elastic lines.
\end{abstract}


\author[1,2]{Antoine Sanner}
\ead{antoine.sanner@imtek.uni-freiburg.de}

\author[1,2]{Lars Pastewka}
\ead{lars.pastewka@imtek.uni-freiburg.de}

\address[1]{Department of Microsystems Engineering, University of Freiburg, Georges-K\"ohler-Allee 103, 79110 Freiburg, Germany}
\address[2]{Cluster of Excellence livMatS, Freiburg Center for Interactive Materials and Bioinspired Technologies, University of Freiburg, Georges-K\"ohler-Allee 105, 79110 Freiburg, Germany}

\date{\today}

\end{frontmatter}

\section{Introduction}

Soft materials stick to rough surfaces if the elastic energy needed to conform to surface
roughness is small compared to the surface energy gained by contact \citep{fuller_effect_1975, briggs_effect_1977, persson_effect_2001}.
In an indentation experiment with such materials,
the force needed to break the contact is often higher than the force measured during indentation 
\citep{chen_molecular_1991
}.
This observation contradicts expectations from classical theories on smooth surfaces \citep{johnson_surface_1971} and theories for rough contacts that assume the contact follows thermodynamic equilibrium
\citep{persson_effect_2001, persson_adhesion_2002-1,persson_adhesion_2002}.
This \emph{adhesion hysteresis} can be caused by several mechanisms like viscoelasticity,
 molecular rearrangements and elastic instabilities. Recent work indicates that in some cases, elastic instabilities triggered by surface roughness play the dominant role \citep{kesari_role_2010, dalvi_linking_2019}.
Details of how roughness gives rise to elastic instabilities remain poorly understood, mainly
because we lack theoretical models accounting for realistic surface roughness.

In this paper, we discuss how to describe the contact of elastic spheres with chemical heterogeneity, local fluctuations in the work of adhesion, using a first-order crack-perturbation model. Understanding the hysteresis caused by quenched disorder in the work of adhesion is a first step towards a general model for geometrically rough surfaces.
Our model will permit efficient numerical simulations of realistic system sizes
and clarify the link between adhesive contact mechanics and classic theories on pinning of elastic lines.

Depending on the compliance of the elastic material, hysteresis arises from disconnected patches snapping in and out of contact
or depinning instabilities in the motion of the contact perimeter, the crack front.
Brute-force numerical methods like the boundary element method (BEM) can capture these different regimes
of instabilities \citep{medina_numerical_2014,dapp_contact_2015, carbone_loading-unloading_2015,deng_molecular_2017,wang_modeling_2021}.
However, for soft materials it is challenging to sufficiently discretize the adhesive neck (or crack tip)
and at the same time include a representative amount of surface roughness \citep{wang_modeling_2021}. Insufficient discretization in boundary element models can additionally lead to ``lattice trapping'', that causes artificial pinning of the crack front alike physical lattice trapping in atomic crystal \citep{thomson_lattice_1971}.

More coarse-grained models either concentrate on the two opposite limits of very low contact fraction or full contact.
Asperity models describe surface roughness as a set of spherical peaks with random heights \citep{greenwood_contact_1966,fuller_effect_1975},
that dissipate energy during snap in and out of contact instabilities
\citep{zappone_role_2007,
wei_effects_2010,
greenwood_reflections_2017,%
deng_depth-dependent_2019,
violano_roughness-induced_2021
}.
This type of approximation breaks down when asperities coalesce and a larger fraction of the surface comes into contact.
In the limit of small roughness and low stiffness, the contact area is simply connected.
In that limit, the surface roughness causes energy barriers that pin the crack front
and the crack front dissipates energy during depinning instabilities.
This phenomenon has to date only been studied on one-dimensional roughness \citep{guduru_detachment_2007-1,guduru_detachment_2007, kesari_role_2010, kesari_effective_2011, carbone_loading-unloading_2015}.

The pinning of a crack front by quenched disorder in the work of adhesion (or equivalently, the fracture toughness)
is better understood
\citep{
gao_first-order_1989,
schmittbuhl_interfacial_1995,
xia_toughening_2012,
demery_microstructural_2014,
xia_adhesion_2015,
chopin_morphology_2015,
ponson_statistical_2016,
lebihain_effective_2021
}.
An essential step towards this understanding was Rice's description of a semi-infinite crack as an elastic line
with long-range elasticity \citep{rice_first-order_1985}.
This equation belongs to the class of elastic interfaces pinned by a random field and the current understanding on fracture of heterogeneous media benefited from works in other fields \citep{
larkin_pinning_1979,
fisher_threshold_1983,
robbins_contact_1987,
middleton_asymptotic_1992,
amaral_scaling_1995,
zhou_critical_2000,
rosso_roughness_2002,
tanguy_weak_2004,
rosso_numerical_2007}.

Rice's first-order perturbation
was applied to several geometries with finite sizes \citep{lazarus_perturbation_2011,patinet_finite_2013}.
To our knowledge, this perturbation approach was never applied to the contact of spheres or similar indenters,
except recently by \citet{argatov_controlling_2021} to investigate the effect of indenter ellipticity on the pull-off force.
Crack perturbation is a promising approach
to understand how surface roughness affects adhesion hysteresis in indentation experiments where the contact
area is simply connected. The first step towards this end is to understand how the adhesion hysteresis depends on
the geometry and strength of quenched disorder in the work of adhesion.

\begin{figure}
	\centering
	\includegraphics[width=\textwidth]{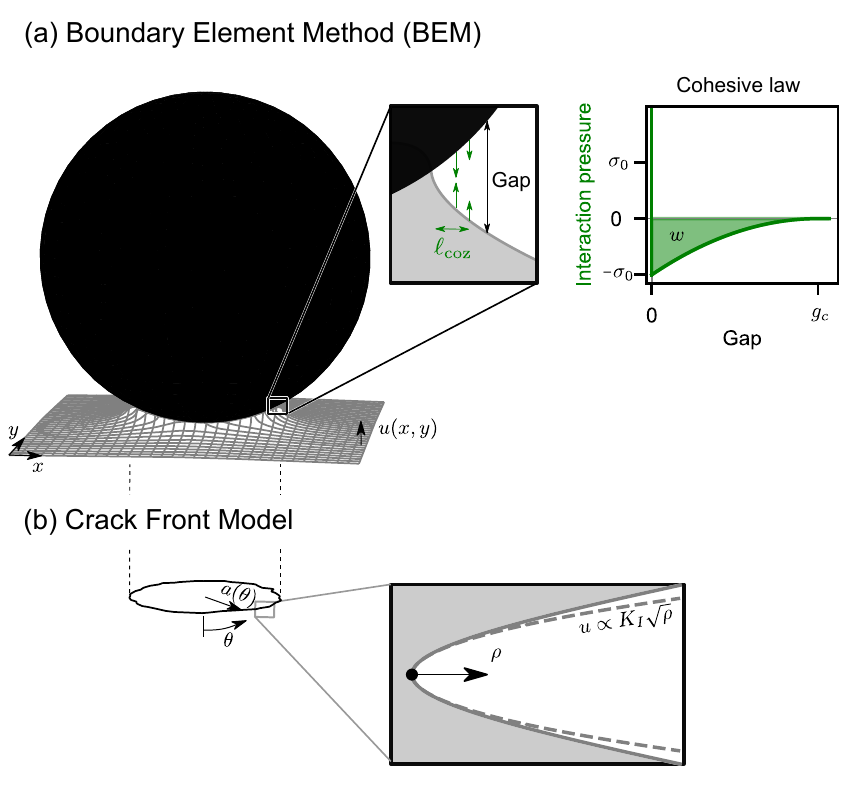}
	\caption{
	Schematic illustration of the two types of models used in this work to describe the contact of a sphere.
	(a) In the boundary element method (BEM), the surface displacements $u(x,y)$ are computed using the Green's function of the elastic half-space
	and
	adhesive attractions are modeled using a cohesive law that relates the interaction pressure to the gap between the two surfaces. The integral over the cohesive law gives the work of adhesion $w$. Spatial heterogeneity is encoded into the cohesive law.
	(b) In the crack-front model, the state of the nearly circular contact is described by the contact radius as a function of angle $a(\theta)$.
	The heterogeneous work of adhesion distorts the shape of the crack front.
	The elastic response of the contact line is approximated using the first-order variation of the stress intensity
	factor with respect to the contact shape developed by \cite{gao_nearly_1987}.
	}
	\label{fig:conceptfigure}
\end{figure}

In this paper, we discuss how to describe the contact of spheres
with fluctuating work of adhesion as a perturbation from the homogeneous, perfectly circular contact, as described by the Johnson-Kendall-Roberts (JKR) theory \citep{johnson_surface_1971,barthel_adhesive_2008}.
The perturbation relies on the first functional derivative of the stress intensity factor derived by~\cite{gao_nearly_1987}.
We describe below that there is no unique way to extrapolate the elastic response of the elastic line.
In the past either the stress intensity factor \citep{gao_first-order_1989, fares_crack_1989}
or the energy release rate \citep{bonamy_crackling_2008, ponson_crack_2010, patinet_finite_2013, chopin_morphology_2015, ponson_statistical_2016}
 was linearly extrapolated. We here propose a third equation that ensures that the crack front possess an elastic potential and therefore constitutes a variational approach to crack perturbation.
In order to validate our model and to discriminate between different variants of the perturbation,
we compare the crack-perturbation models to finely discretized BEM simulations.
Figure~\ref{fig:conceptfigure} illustrates schematically the two types of models used in this paper.

\section{Problem definition}
\label{sec:problem_definition}

We consider the contact of a sphere (to be exact, a paraboloid) adhering an elastic half-space at a fixed rigid body penetration $\Delta$.
This case can be mapped to the contact of two spheres with the same composite radius $R$
and contact modulus $E^\prime$ \citep{johnson_contact_1985}.
When only one half-space deforms, $E^\prime=E/(1-\nu^2)$, where $E$ is Young's modulus and $\nu$ is Poisson's ratio.
Fracture mechanics typically considers the contact of two elastic half-spaces where $E^\prime=E / 2 (1-\nu^2)$.
We assume the contact is frictionless and consider only vertical displacements.

The equilibrium contact area minimizes the total energy $\Pi = \Pi_\mathrm{mech} + \Pi_\mathrm{surf}$.
The mechanical energy $\Pi_\mathrm{mech} $ contains the elastic strain energy $U_\mathrm{el}$ and the potential of external forces.
The surface energy $\Pi_\mathrm{surf}$ results from adhesive interactions.
We consider the limit of infinitesimally short interaction range (JKR-limit), where
\begin{equation}
\Pi_{\mathrm{surf}} = \int \limits_{A_c} \dif x \dif y\ w(x, y).
\end{equation}
$A_c$ is the contact area and the work of adhesion $w(x,y)$ depends on the position $(x,y)$.
Note that the work of adhesion is the same for a receding crack (indentation) as for an opening crack (retraction),
i.e. there is no \emph{intrinsic} hysteresis in the work of adhesion.

Once nondimensionalized using distinct vertical and lateral length units, the JKR contact is parameter free \citep{muller_influence_1980, maugis_contact_2010, muser_single-asperity_2014}.
We present our numerical results in nondimensional units \citep{maugis_contact_2010}
and indicate by an asterisk$^*$ when quantities are normalized.
Specifically,
lengths along the surface of the half-space (e.g., the contact radius) are normalized by $(3\pi w_m R^2 / 4 E^\prime )^{1/3}$,
lengths in vertical direction (e.g., displacements) by $(9\pi^2 w_m^2 R^2 / 16 {E^\prime}^2)^{1/3}$
and normal forces by $\pi w_m R$.
$w_m$ is the median work of adhesion.
The equations are in dimensional form but can be nondimensionalized by substituting $R=1$, $w_m = 1/\pi$ and $E^\prime = 3 / 4$.

\section{Reference model}
\label{sec:BEM}

We used the BEM to validate the crack-front models.
BEM computes the surface displacements $u(x, y)$ (Fig.~\ref{fig:conceptfigure}a) minimizing the total energy $\Pi$ at prescribed rigid body penetration $\Delta$.
This method makes no assumptions on the contact morphology, allowing for holes and disconnected contact islands,
but it is computationally too expensive to properly discretize contacts with small heterogeneities.

The mechanical energy $\Pi_\mathrm{mech} = U_\mathrm{el}$ and its gradients are computed using the Green's function
for pressures that are constant on each pixel~\citep{love_stress_1929, johnson_contact_1985}.
While the Green's function is nonperiodic, we accelerated calculations with a
fast Fourier transform~\citep{stanley_fft-based_1997,campana_practical_2006,pastewka_seamless_2012}
by introducing a padding region that decouples periodic images~\citep{hockney_potential_1970, liu_versatile_2000, pastewka_contact_2016}.

The surface energy needs to be regularized using an interaction potential $\phi$ between the surfaces (cohesive law),
\begin{equation}
\Pi_{\mathrm{surf}} = \int \limits_{-\infty}^{+\infty} \dif x \dif y \ \phi\left(w(x, y),  g(x,y)\right),
\end{equation}
where $w(x,y)$ is the (spatially varying) work of adhesion and $g(x,y)$ is the gap between the surfaces.
The results converge to the JKR limit for small pixel size and interaction range, i.e. high Tabor parameter $\mu_\mathrm{T}$
 \citep{muller_influence_1980, maugis_adhesion_1992, greenwood_adhesion_1997, muser_single-asperity_2014},
but the point of instability where the two interfaces jump into contact converges particularly slowly with the interaction range \citep{wu_jump--contact_2010, ciavarella_effect_2017}.
\cite{wang_modeling_2021} could slightly retard the premature jump into contact by using a potential with a cutoff. Our choice of cohesive law,
\begin{equation}
\phi(w, g) =
\begin{cases}
- w \left(1  - \frac{g}{g_c}\right)^3, & g < g_c
\\
0, & \text{else}
\end{cases},  
\label{eq:pot}
\end{equation}
is based on the same observation. Equation~\eqref{eq:pot} describes a cubic potential where $w$ is the work of adhesion, $g$ is the gap and $g_c$ the cutoff distance.
The derivative of $\phi$ (interaction pressure) is illustrated in Fig.~\ref{fig:conceptfigure}a.
The maximum attractive stress $\sigma_{0} = 3 w / g_c$ occurs at the perimeter of the contact area, where $g = 0$.
To model (chemical) heterogeneity, the interaction parameters are different in each pixel.
We keep $g_c$ constant, so that $\sigma_0$ and $w$ vary proportionally from pixel to pixel.

The surfaces repel each other with a hard-wall potential, which is implemented as inequality constraints ($g(x,y)\geq 0$) in the minimization algorithms.
We used the constrained quasi-newton algorithm L-BFGS-B~\citep{byrd_limited_1995}
and a constrained conjugate gradient algorithm~\citep{polonsky_numerical_1999,vollebregt_bound-constrained_2014,bugnicourt_fft-based_2018}
to minimize the energy.
The latter algorithm is parallelized with the message passing interface and was used to produce precise reference solutions for the contact shape,
but could not be used in the presence of instabilities.
Note that these minimization algorithms require that the interaction potential has a continuous second derivative, justifying our choice of a third-order polynomial.
BEM with hard-wall repulsion and finite-ranged attraction \citep{muser_single-asperity_2014,muser_dimensionless_2016,muser_meeting_2017,wang_gauging_2017,
bazrafshan_numerical_2017,rey_normal_2017,bugnicourt_fft-based_2018,monti_distribution_2021} or soft (Lennard-Jones type) repulsion
\citep{greenwood_adhesion_1997,feng_contact_2000,
wu_jump--contact_2010,medina_numerical_2014,
pastewka_contact_2014,pastewka_contact_2016,persson_theory_2014,monti_effect_2019,
ghanbarzadeh_deterministic_2020,wang_modeling_2021} have been used in the past to study the adhesion of spheres and rough surfaces.
During retraction, these models are similar to fiber-bundle models of quasi-brittle fracture
~\citep{batrouni_heterogeneous_2002, schmittbuhl_roughness_2003, stormo_onset_2012, gjerden_universality_2013, gjerden_local_2014}
and threshold-force models~\citep{pohrt_adhesive_2015,hulikal_relation_2017,li_adhesive_2019}.

The finite interaction range introduces a cohesive zone around the contact perimeter in which the surfaces attract each other.
This cohesive zone needs to be small in comparison to the scale of the heterogeneity~\citep{chen_apparent_2008} in order to mimic
a JKR-like contact.
We estimate the width of the cohesive zone using a Dugdale model and assume it
is small in comparison to the contact radius \citep{maugis_contact_2010}: 
\begin{equation}\label{eq:cohesive_zone_size}
\ell_\text{coz} = \frac{\pi K_I^2 }{8 \sigma_0^2} = \frac{\pi E^\prime w }{4 \sigma_0^2}
\end{equation}
For uniform work of adhesion, \cite{muser_single-asperity_2014} showed that the (nondimensionalized) contact radius difference
with the JKR limit is asymptotically $\sim \ell_\mathrm{coz}^* \sim \mu_\mathrm{T}^{-2}$
, where $\ell_\mathrm{coz}^* = \ell_\mathrm{coz} / (3\pi w_m R^2 / 4 E^\prime )^{1/3}$ is the nondimensionalized cohesive zone width,
see section~\ref{sec:problem_definition}.
We use $\ell^*_\text{coz}$ as a proxy for interaction range and Tabor parameter $\mu_\mathrm{T}$ \citep{tabor_surface_1977,muller_influence_1980,maugis_adhesion_1992}.
We chose the pixel size $\ell_\mathrm{pix}$ small enough so that further grid refinement affects the contact radius and the force
less than decreasing the interaction range. 

\section{Crack-front model}

The crack-front model is a first-order perturbation from the axisymmetric JKR-contact of a sphere \citep{johnson_surface_1971,barthel_adhesive_2008}
using Rice's weight function theory \citep{rice_first-order_1985, gao_nearly_1987, wei_weight_1989}.

\subsection{Axisymmetric contact: The JKR model}
\label{sec:JKR}

We briefly review the JKR model of the adhesion of a sphere against an elastic half-space.
JKR described the contact by the balance of elastic and surface energy of a circular crack \citep{griffith_vi._1921}.
At equilibrium, the contact radius $a$ is such that the increment of mechanical energy equals the increment of surface energy:
\begin{equation}\label{eq:jkr_err_balance}
\frac{\partial \Pi_\mathrm{mech}(a, \Delta)}{\partial a} = 2 \pi a w(a)
\end{equation}
$\Pi_\mathrm{mech}$ is the mechanical potential and $w$ is the work of adhesion at the perimeter of the contact.
In this paper, we consider only boundary conditions with fixed rigid body penetration $\Delta$, where $\Pi_\mathrm{mech}$ is the elastic energy $U_\mathrm{el}.$

The elastic energy release rate,
\begin{equation}\label{eq:definition_G}
G(a, \Delta) = \frac{1}{2 \pi a}\frac{\partial \Pi_\mathrm{mech}(a, \Delta)}{\partial a},
\end{equation}
can be related to the intensity $K$ of the singularity of the stress field near the crack tip \citep{irwin_analysis_1957}:
\begin{equation}\label{eq:Irwin}
G = \frac{K^2}{2 E^\prime}.
\end{equation}
JKR obtained the stress intensity factor $K_\text{J}$ for sphere on flat contact by superposing contact pressures of the Hertzian sphere \citep{hertz_ueber_1881}
and of the circular flat punch \citep{sneddon_boussinesqs_1946, maugis_contact_2010}:
\begin{equation}\label{eq:jkr_sif}
K_\text{J}(a, \Delta) =\left(\frac{a^2}{R} - \Delta\right) \frac{E^\prime}{\sqrt{\pi a}}
\end{equation}
Inserting Eqs.~\eqref{eq:definition_G}~to~\eqref{eq:jkr_sif} into Eq.~\eqref{eq:jkr_err_balance}
yields an equilibrium equation for the contact radius as a function of the rigid body penetration.
The normal force follows from the same superposition \citep{johnson_surface_1971,maugis_contact_2010}:
\begin{equation}\label{eq:jkr_force}
F_\text{J}(a, \Delta ) = \frac{4 E^\prime}{3 R} a^3 + 2aE^\prime \left(\Delta - \frac{a^2 }{R}\right)
\end{equation}
The increment in surface energy is the integral of the work of adhesion within the area swept by the contact line.
The JKR theory applies to any axisymmetric  work of adhesion, where $w$ depends on the radius in Eq. \eqref{eq:jkr_err_balance}.

\subsection{Equilibrium condition for non-circular contacts}

\label{sec:contact_line}

We now consider the perimeter of a nearly circular contact.
The function $a(s)$ represents the planar distance between the point at arc length $s$
on the contact perimeter (or crack front) and the tip of the sphere.
The mechanical potential is now a \emph{functional} of the contact radius $a$.
The elastic energy release rate $G(s)= \delta \Pi_\mathrm{mech}/\delta a (s)$ is the
functional derivative of the mechanical potential $\Pi_\mathrm{mech}$ with respect to the contact radius:
\begin{equation}
\delta \Pi_\mathrm{mech}([\delta a]) = \oint \dif s \ \delta a(s) G(s)
\end{equation}
We omit the penetration $\Delta$ and the contact shape $a$ from the arguments for brevity.
Note that $\dif s \ \delta a(s)$ is an infinitesimal surface increment, so that $G$ is an energy per surface area, as in Eq.~\eqref{eq:definition_G}. Here an below we use square brackets $[\cdot]$ to indicate a functional dependence.

The stationarity of the total potential, i.e. the Griffith criterion, requires that the mechanical energy release rate at point $s$ on
the contact perimeter equals the local work of adhesion at the point $(a(s), s)$ on the heterogeneous plane:
\begin{equation}\label{eq:Griffith_local}
G(s) = w(a(s), s)
\end{equation}
$G$ is available in closed form only for simple crack and contact shapes.
Rice's weight function theory allows us to compute efficient approximations of $G$ when perturbed from a reference
configuration.

\begin{figure}
\centering
\includegraphics[width=0.8\textwidth]{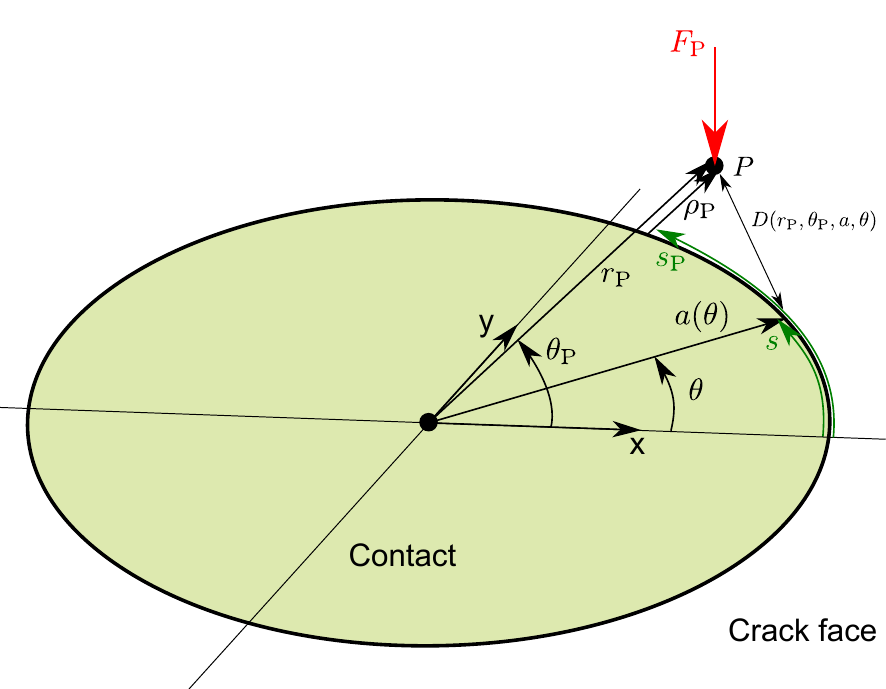}
\caption{The crack-face weight function $k$
is central in the derivation of the first-order variation of the stress intensity factor.
$k_\mathrm{J}(a, r_\mathrm{P}, \theta_\mathrm{P}, \theta)$
corresponds to the stress intensity factor at angle $\theta$
when a unit point force $F_\mathrm{P}$ is applied on the face of a circular external crack at
$r_\mathrm{P}, \theta_\mathrm{P}$.
The origin of the coordinate system is aligned with the tip of the spherical indenter.
}
\label{fig:schema_coordinates}
\end{figure}

Below we will parameterize the crack front using the angle $\theta$ instead of the arc length $s$, see Fig.~\ref{fig:schema_coordinates}.
In this new parameterization,
\begin{equation}
\delta \Pi_\mathrm{mech}([\delta a]) = \oint \dif s \ G(s) \delta a(s) = \int \limits_0^{2\pi} \dif \theta \  a(\theta) G(\theta) \delta a(\theta)
,
\end{equation}
so that the functional derivative of the potential
\begin{equation}\label{eq:dU_da_theta}
\frac{\delta \Pi_\mathrm{mech}}{\delta a(\theta)} = a(\theta) G(\theta)
\end{equation}
instead of $G(\theta)$.

\subsection{First-order variation of the stress intensity factor}

Rice showed that the sensitivity of the stress intensity factor to the contact shape $\delta K / \delta a$ is linked to the crack-face weight function.
The crack-face weight function $k([a]; P, s)$ is the stress intensity factor caused by a unit force at point $P$ outside the contact area
and is known for simple geometries \citep{tada_stress_2000}.
While Rice's first variation of $K$ was almost exclusively applied to flat-on-flat contact geometries (planar crack problems),
$\delta K / \delta a$ is unaffected by the presence of the Hertzian displacements in the contact area \citep{rice_three-dimensional_1985, borodachev_contact_1991}.
To make explicit that the first-order variation of $K$ for circular connections \citep{gao_nearly_1987}
applies also to the contact of spheres, we review the main steps leading to $\delta K / \delta a$.

We apply a force $F_\mathrm{P}$ on the elastic half-space at point $P$ outside the contact area (on the crack face).
This point force causes the displacement $u_\mathrm{P}$ at $P$.
At fixed rigid body penetration $\Delta$ and fixed force $F_\mathrm{P}$, the mechanical potential is the Legendre transform of the elastic strain energy $U_\mathrm{el}$,
\begin{equation}
\Pi_\mathrm{mech}(F_\mathrm{P}) = \min_{u_\mathrm{P}} \left\{ U_\mathrm{el}(u_\mathrm{P}) - u_\mathrm{P} F_\mathrm{P} \right\}
,
\end{equation}
with derivatives
\begin{equation}
\left. \frac{\partial \Pi_\mathrm{mech}}{\partial F_\mathrm{P}}\right|_{a} = - u_\mathrm{P}
\qquad\text{and}\qquad
\left. \frac{\delta \Pi_\mathrm{mech}}{\delta a(s)}\right|_{F_\mathrm{P}} = \left. \frac{\delta U_\mathrm{el}}{\delta a(s)}\right|_{u_\mathrm{P}} = G(s)
.
\end{equation}
$U_\mathrm{el}$ is the strain energy and $-u_\mathrm{P} F_\mathrm{P}$ is the potential of the applied force $F_\mathrm{P}$.
Because the second derivatives of $\Pi_\mathrm{mech}$ are continuous, the order of differentiation does not matter (Schwartz's theorem):
\begin{equation}\label{eq:du_G_symmetry}
\frac{\delta u_\mathrm{P}}{\delta a(s)} = - \frac{\partial G(s)}{\partial F_\mathrm{P}}
.
\end{equation}
From Eq.~\eqref{eq:Irwin} and $\partial K / \partial F_\mathrm{P} = k([a]; P, s)$ we obtain
\begin{equation}\label{eq:du_Kk}
\frac{\delta u_\mathrm{P}}{\delta a(s)}
 =  - \frac{1}{ E^\prime} K(s) k([a]; P, s).
\end{equation}
Setting $F_\mathrm{P}=0$ (for arbitrary $P$) we see that the crack-face weight function $k$ describes
how the crack faces deform when the crack front moves.

Knowing that the displacements near the crack tip correspond to the stress intensity factor,
we can now deduce how the stress intensity factor is affected by a perturbation of the contact radius $\delta a$.
At the perpendicular distance $\rho_\mathrm{P}$ away from the point $s_\mathrm{P}$ on the crack tip,
\begin{equation}\label{eq:u_K_expansion}
u(\rho_\mathrm{P}, s_\mathrm{P}) = \frac{4}{E^\prime} \sqrt{\frac{\rho_\mathrm{P}}{2\pi}} K(s_\mathrm{P}) + \mathcal{O}(\rho_\mathrm{P}^{3/2}).
\end{equation}
From Eq.~\eqref{eq:u_K_expansion}, the perturbation of the stress intensity factor becomes
\begin{equation}\label{eq:dK_from_du_limit_da0}
\begin{split}
\delta K(s_\mathrm{P})
&= \lim \limits_{\rho_\mathrm{P} \to 0} \frac{E^\prime}{4} \sqrt{\frac{2\pi}{\rho_\mathrm{P}}}
\oint ds \ \delta a(s) \frac{\delta u(\rho_\mathrm{P}, s_\mathrm{P})}{\delta a(s)}
\\
&=
-
\lim \limits_{\rho_\mathrm{P} \to 0} \sqrt{\frac{\pi}{8 \rho_\mathrm{P}}}
\oint ds \ \delta a(s) K(s) k([a];\rho_\mathrm{P}, s_\mathrm{P}, s)
,
\end{split}
\end{equation}
for $\delta a$ restricted to $\delta a(s_\mathrm{P}) = 0$.
Rice introduced this restriction to ensure that
$\delta \rho_\mathrm{P} = 0$.
As a consequence, Eq.~\eqref{eq:dK_from_du_limit_da0} is not a first order variation because $\delta a$ is not arbitrary.
This restriction will be relaxed later by expanding the circular reference configuration by $\delta a(s_\mathrm{P})$.

When the contact perimeter is a circle  ($a(s) = \mathrm{const.}$),
we know $K$ from \cite{johnson_surface_1971} and $k$ from Galin \citep{gladwell_contact_2008},
enabling us to evaluate Eq.~\eqref{eq:dK_from_du_limit_da0}.
For convenience, we now parameterize the position along the contact line with the angle $\theta$ instead of the arc length $s$.

The crack-face weight function for the JKR contact is the same as for a circular connection.
For a circular connection with radius $a$ and fixed displacements at infinity \citep{gao_nearly_1987, wei_weight_1989, gladwell_contact_2008},
\begin{equation}\label{eq:circular_weight_function}
k_\text{J}(a, r_\mathrm{P}, \theta_\mathrm{P},  \theta) = \frac{\sqrt{(r_\mathrm{P}^2 - a^2) / (a\pi^3) }}{D^2(r_\mathrm{P}, \theta_\mathrm{P}, a, \theta)}
,
\end{equation}
where $D^2(r_\mathrm{P}, \theta_\mathrm{P}, a, \theta) = a^2 + r_\mathrm{P}^2 - 2 a r_\mathrm{P} \cos(\theta_\mathrm{P} - \theta)$ is the square of the distance between the point of
application of the force on the crack face, $P = (r_\mathrm{P}, \theta_\mathrm{P})$, and the point on the crack-front
where the stress intensity factor is calculated, $\theta$. Note that $k_\text{J}$ is not a functional of the whole radius, since in JKR the underlying contact area is a circle with radius $a$.

First applying a uniform perturbation of magnitude $\delta a(\theta_\mathrm{P})$ to the circular reference contact (radius $\tilde a_0$)
and then using $K=K_\text{J}$ and inserting Eq.~\eqref{eq:circular_weight_function} into Eq.~\eqref{eq:dK_from_du_limit_da0}
yields the first order variation of $K(\theta_\mathrm{P})$ at $a=\tilde a_0 = \mathrm{const.}$:
\begin{align}\label{eq:dK_1}
\delta K (\theta_\mathrm{P})
=&
\frac{\partial K_\text{J}(\tilde a_0)}{\partial \tilde a_0} \delta a(\theta_\mathrm{P})
- \frac{K_\text{J}(\tilde a_0)}{8\pi} \
\text{PV}\int  \limits_{0}^{2\pi} d\theta \
\frac{\tilde a_0 (\delta a(\theta) - \delta a (\theta_\mathrm{P}))}{D^2(r_\mathrm{P}, \theta_\mathrm{P}, \tilde a_0, \theta)}
\end{align}
We point to~\cite{gao_nearly_1987,gao_somewhat_1987} for details on how the limit $\rho_\mathrm{P} \to 0$ leads to the Cauchy principal value integral $\text{PV}\int$.

We now recast Eq.~\eqref{eq:dK_1} using the Fourier series of $a$. Inserting
\begin{equation}
a(\theta) = \sum \limits_{n \in \mathbb{Z}} \tilde{a}_n e^{i n \theta}
\label{eq:fourierinv}
\end{equation}
into Eq.~\eqref{eq:dK_1} yields
\begin{equation}\label{eq:circular_dK_of_da_fourier}
\delta K (\theta_\text{P})
=
\frac{\partial K_\text{J}(\tilde a_0)}{\partial \tilde a_0} \delta a(\theta_\text{P})
+
\frac{K_\text{J}(\tilde a_0)}{\tilde a_0}
\sum \limits_{n \in \mathbb{Z}^{\backslash \{0\}}}  \frac{|n|}{2} \  \delta \tilde a_n \ e^{i n \theta_\text{P}}.
\end{equation}
From the inverse of Eq.~\eqref{eq:fourierinv},
\begin{equation}
    \tilde a_n = \frac{1}{2\pi}\int_0^{2\pi}d\theta \  e^{- i n \theta }  a(\theta),
\end{equation}
we obtain $\delta \tilde{a}_n/\delta a(\theta) = \exp(-i n \theta)/2\pi$. The
functional derivative of $K$ then becomes
\begin{equation}\label{eq:dK_da}
\frac{\delta K}{\delta a(\theta)}
= \frac{\partial K_\text{J}(\tilde a_0)}{\partial \tilde a_0} \delta(\theta - \theta_\text{P})
+
\frac{K_\text{J}(\tilde a_0)}{2 \pi \tilde a_0}
\sum \limits_{n \in \mathbb{N}}  |n|
\cos n(\theta_\text{P} - \theta),
\end{equation}
which depends only on $|\theta - \theta_\text{P}|$.
The translational invariance reflects the axial symmetry of the unperturbed state.
As will be discussed in a few paragraphs, $\theta$ and $\theta_\text{P}$ commute because $\delta K / \delta a(\theta)$ is related to
the second derivative of the energy.

The key step leading to $\delta K / \delta a(\theta)$ was recognizing the symmetry of the derivatives of the elastic energy, Eq.~\eqref{eq:du_G_symmetry},
that follows from the smoothness of the elastic energy.
Hence, Eq.~\eqref{eq:dK_da} not only applies to flat indenters
\citep{rice_first-order_1985, gao_nearly_1987, gao_somewhat_1987, wei_weight_1989},
but also to contacts against spheres or any smooth indenters.
All the relevant information of the axisymmetric contact geometry is in $K_\mathrm{J}$ and its derivatives.
Knowing the first order variation of $K$ when the contact is perfectly circular allows us to approximate
the equilibrium Eq.~\eqref{eq:Griffith_local} to first order in the perturbation from circularity.

\subsection{Quadratic approximation of the deformation energy}
The straightforward approach to construct a first order model is to linearly extrapolate either $G$ or $K$.
However, these approximations do not conserve the variational property of $a G$, i.e. that it is the gradient of a potential, Eq.~\eqref{eq:dU_da_theta}.
We now guess a quadratic approximation for the elastic energy
and verify a posteriori that it possesses the exact first and second derivatives at $a(\theta) = \tilde a_0$. Our guess for the elastic energy is
\begin{equation}
\label{eq:circular_U}
U_\text{el} = \frac{1}{2\pi}\int \limits_0^{2 \pi} d\theta \ U_\text{J}(a(\theta)) + \pi G_\text{J}(\tilde a_0) \sum_n|n||\tilde a_n|^2
\end{equation}
with first derivative
\begin{equation}
\label{eq:circular_dU_da}
\frac{\delta U_\text{el}}{\delta a(\theta)}=
\frac{1}{2\pi}  \frac{\partial U_\text{J}(a(\theta))}{\partial a(\theta)}
+ G_\text{J}(\tilde a_0)   \sum_n |n| \tilde a_n e^{i n \theta}
+ \frac{1}{2} \frac{\partial G_\text{J}(\tilde a_0) }{\partial \tilde a_0}  \sum_n |n| |\tilde a_n|^2
\end{equation}
and second derivative
\begin{equation}
\label{eq:circular_d2U_da2}
\begin{split}
\frac{\delta^2 U_\text{el}}{\delta a(\theta) \delta a(\theta_\text{P})} =&
 \frac{1}{2\pi} \frac{\partial^2 U_\text{J}(a(\theta))}{\partial a^2(\theta)} \delta(\theta - \theta_\text{P})
+ \frac{1}{2\pi} G_\text{J}(\tilde a_0) \sum_n |n| e^{i n (\theta - \theta_\text{P})}
\\&
+ \frac{1}{2\pi}  \frac{\partial G_\text{J}(\tilde a_0) }{\partial \tilde a_0}  \sum_n |n| \tilde a_n (e^{i n \theta} + e^{i n \theta_\text{P}})
+ \frac{1}{4\pi} \frac{\partial^2 G_\text{J}(\tilde a_0) }{\partial \tilde a_0^2}  \sum_n |n| |\tilde a_n|^2.
\end{split}
\end{equation}
$U_\text{J}$ is the elastic energy in the perfectly circular contact \citep{johnson_surface_1971}
and $\partial U_\text{J}/\partial a = 2\pi a G_\text{J}(a)$.
It is straightforward to show that Eq.~\eqref{eq:circular_d2U_da2} for $a(\theta)=\tilde{a}_0$ gives Eq.~\eqref{eq:dK_da}.
We determine the contact shape by solving (see also Eq.~\eqref{eq:dU_da_theta})
\begin{equation}\label{eq:CF-E_equilibrium}
\frac{\delta U_\text{el}}{\delta a(\theta)} - w(\theta, a(\theta)) a(\theta) = 0
\end{equation}
using a minimization algorithm (see section \ref{sec:numerical_implementation}).
We will call this model CF-E when presenting our results.

The normal force applied on the indenter follows directly from our approximation of the energy:
\begin{equation}\label{eq:normal_force_from_energy}
F = \frac{\partial U_\mathrm{el}(\Delta, a)}{\partial \Delta}
.
\end{equation}
Using Eq. \eqref{eq:circular_U},
\begin{equation}\label{eq:circular_F}
F(a, \Delta)
=
\frac{1}{2\pi}\int \limits_{0}^{2\pi} d\theta \ F_\text{J}(a(\theta), \Delta)
+ \pi \frac{\partial G_\text{J}(\tilde a_0, \Delta)}{\partial \Delta} \sum_n |n||\tilde a_n|^2.
\end{equation}

\subsection{Other first order approximations}
\label{sec:other_CF_models}

As a benchmark we also consider two common first order approximations of the crack front.
The first approximation linearly extrapolates the stress intensity factor and solves for
$K=K_\mathrm{c} = \sqrt{2 E^\prime w}$~\citep{gao_first-order_1989, fares_crack_1989}:
\begin{equation}\label{eq:K-lin}
K(\theta) = K_\mathrm{J}(\tilde a_0)
+ \frac{\partial K_\mathrm{J}(\tilde a_0)}{\partial \tilde a_0} \left(a(\theta) - \tilde a_0\right)
+ \frac{K_\mathrm{J}(\tilde a_0)}{\tilde a_0}
\sum \limits_{n \in \mathbf{Z}}  \frac{|n|}{2} \tilde a_n e^{i n \theta}
.
\end{equation}
We will refer to this model as CF-K.

The second approximation linearly extrapolates the energy release rate and solves for $G=w$~\citep{bonamy_crackling_2008,chopin_morphology_2015,ponson_statistical_2016}:
\begin{equation}\label{eq:G-lin}
G(\theta) = G_\mathrm{J}(\tilde a_0)
+ \frac{\partial G_\mathrm{J}(\tilde a_0)}{\partial \tilde a_0} (a(\theta) - \tilde a_0)
+ \frac{G_\mathrm{J}(\tilde a_0)}{\tilde a_0} \sum_{n \in \mathbf{Z}} |n| e^{i n \theta} \tilde a_n
.
\end{equation}
We will refer to this model as CF-G.

An alternative first order extrapolation,
$$
G(\theta) = G_\mathrm{J}(a(\theta)) \left( 1
+ \frac{1}{ a(\theta)} \sum_{n \in \mathbf{Z}} |n| e^{i n \theta} \tilde a_n
\right),
$$
should in principle be more accurate, because $G_\mathrm{J}(a(\theta))$ is not linearized.
However we observed that using the latter equation does not improve the prediction of the normal force and Eq.~\eqref{eq:G-lin} is better suited
for analytical models.
We did the same observation with the analogous expression for the stress intensity factor.

Because these crack-front models do not possess an elastic energy, we cannot use Eq.~\eqref{eq:normal_force_from_energy}
to compute the normal force.
As an approximation, we neglect the effect of the undulations of the crack front on the normal force and insert the mean
contact radius $\tilde{a}_0$ into the expression for a perfectly circular contact, Eq.~\eqref{eq:jkr_force}.
We will discuss in section~\ref{sec:validate_on_rays} that the crack shape affects the normal force only to second order.

\subsection{Symmetries}
\label{sec:symmetries}

Rice obtained the first order variation of $K$ (Eq.~\eqref{eq:dK_da})
from the symmetry of second derivatives (or the path independence of the work)
with respect to the contact radius and a point force applied on the crack face.
Similarly, the second derivatives with respect to the contact radius are symmetric
\citep{gao_first-order_1989, leblond_second-order_2012, salvadori_weight_2014}:
\begin{equation}\label{eq:functional_schwartz}
\frac{\delta^2 U}{\delta a(\theta) \delta a(\theta_\text{P}) } = \frac{\delta^2 U}{\delta a(\theta_\text{P}) \delta a(\theta) }
\end{equation}
For the energy release rate, it follows from Eq.~\eqref{eq:dU_da_theta} that
$a(\theta) \delta G(\theta) / \delta a (\theta_\text{P})$
should be symmetric.

The linear extrapolation of $K$ (Eq.~\eqref{eq:K-lin}) violates this symmetry.
Linearizing $G$ (Eq.~\eqref{eq:G-lin}) fulfills the symmetry only for translational invariance in the crack propagation direction,
$\partial G_J / \partial a = 0$, i.e. in the case of a semi-infinite crack \citep{leblond_second-order_2012}.

\subsection{Numerical implementation of the crack-front model}
\label{sec:numerical_implementation}

For a fixed penetration $\Delta$, we solve for the equilibrium of energy release rate and work of adhesion
in $N$ collocation points at evenly spaced angles $\theta_j$ (Eq.~\eqref{eq:Griffith_local}).
This turns the functional derivatives presented above into partial derivatives
and $\tilde a_n$ becomes the discrete Fourier transform of the contact radius.
To discretize integrals like $\int_0^{2\pi} d\theta \  U_\mathrm{J}(a(\theta))$,
we take $a$ constant on each angle element of size $\Delta \theta = 2 \pi/N$.

We are interested in stable equilibria,
therefore it is important to use a minimization algorithm and not a root finder.
When we simulate the indentation and retraction process, we use the solution obtained at the previous penetration
as initial guess.
This procedure mimics the quasi-static dynamics of the contact, where the crack front moves to the closest metastable state.

When the crack front moves to the next metastable state through an instability, we need a robust minimization algorithm.
We use a trust-region Newton conjugate-gradient algorithm \citep[Algorithm 7.2]{steihaug_conjugate_1983,nocedal_numerical_2006},
where we \emph{fix}  the radius of the trust region.
Since the work of adhesion field is the source of nonlinearity, the trust radius needs to be slightly smaller than the size of the heterogeneity.
If the minimum contact radius is smaller than this value during a Newton step,
the trust region is further reduced below the contact radius in order to avoid negative contact radii.
Note that trust-region algorithms usually tune the radius according to the
discrepancy between the quadratic subproblem and the actual potential.
We omitted that feature because, as discussed above, some crack-front models do not possess an elastic energy.
An alternative to fixing the trust radius would be to tune it according to gradients instead of energies.

\section{Validation and comparison of the perturbation methods}
\label{sec:validate_on_rays}
\begin{figure}
	\includegraphics[width=\textwidth]{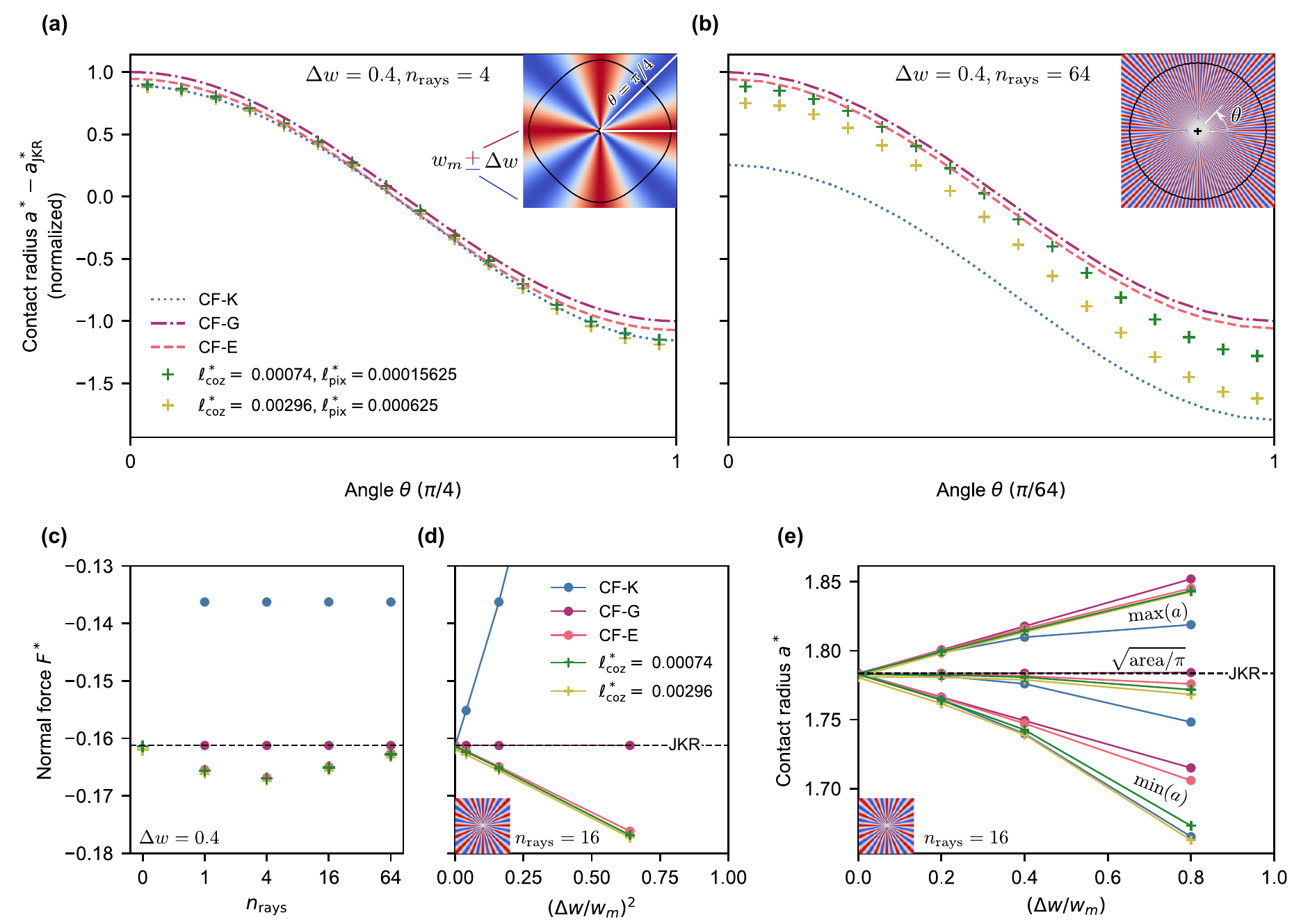}
	\caption{
	Comparison of the different crack-front models with BEM simulations for ray-shaped work of adhesion heterogeneity.
	(a,b)
	Contact radii computed using BEM and the different CF models on the cosinusoidal work of adhesion landscapes illustrated in the
	insets.
	We show the contact radii $a(\theta)$ only over a half period of heterogeneity.
	We used the radii predicted by CF-G (Eq.~\eqref{eq:G-lin}) to normalize the contact radius.
	Since the amplitude of the contact radius is roughly proportional to the wavelength of the heterogeneity,
	the magnification is higher for $n_\mathrm{rays} = 64$ and the effect of the cohesive zone size in the BEM simulations is more apparent.
	(c) Normal force as a function of the number of rays.
	$n_\mathrm{rays}=0$ corresponds to a uniform work of adhesion, where it can be verified that BEM is close
	to the short-range limit (JKR). $F^*$ is the normal force normalized by $\pi w_m R$, as discussed in section~\ref{sec:problem_definition}.
	(d) Normal force as a function of the (squared) amplitude of the work of adhesion heterogeneity, for $n_\mathrm{rays}=16$.
	In the crack-front models CF-K and CF-G, the normal force is computed by inserting the mean contact radius into the JKR equation (Eq.~\eqref{eq:jkr_force}).
	In CF-E, the normal force is computed by taking the derivative of the elastic energy (Eq.~\eqref{eq:circular_F}).
	(e) Contact area, minimum contact radius
	and maximum contact radius in BEM simulations and different crack-front models
	as a function of the amplitude of the work of adhesion heterogeneity.
	}
	\label{fig:convergence_bem_rays}
%
%

\end{figure}

We investigate which first-order crack-perturbation method -- linear extrapolation of $K$ (CF-K, Eq.~\eqref{eq:K-lin}),
linear extrapolation of $G$ (CF-G, Eq.~\eqref{eq:G-lin})
or quadratic extrapolation of the energy (CF-E, Eqs.~\eqref{eq:circular_U}-\eqref{eq:circular_d2U_da2}) -- describes most accurately the contact area and the normal force.
We use BEM simulations with $\ell_\mathrm{coz}^* \simeq 0.0007 $ and pixel size $\ell_\mathrm{pix}^* = 0.00015625$ as a reference, the grid size is $32768\times 32768$ without the padding region.
As a test case, we consider the ray-shaped work of adhesion heterogeneity (see inset to Fig.~\ref{fig:convergence_bem_rays}a)
\begin{equation}\label{eq:rays_heterogeneity}
w(r, \theta) = w_m + \Delta w \cos(n_\text{rays} \ \theta),
\end{equation}
where we vary $n_\mathrm{rays}$ and $\Delta w$.
We fix the value of the penetration to $\Delta^* = 1$.

Figure~\ref{fig:convergence_bem_rays}a,b compares the contact geometry obtained from BEM and the different crack-front models
for $\Delta w = 0.4$ and $4$ rays (Fig.~\ref{fig:convergence_bem_rays}a) and $64$ rays (Fig.~\ref{fig:convergence_bem_rays}b).
To estimate the remaining deviations of BEM from the short-ranged limit, we also show a BEM simulation with a four times larger cohesive zone
($\ell_{\rm coz}^* \simeq 0.0028$) and pixel size ($\ell_\mathrm{pix}^* = 0.000625$, $8192\times 8192$ pixels).
For decreasing $\ell_\mathrm{coz}^*$, the contact radius increases and the normal force decreases
and
when the work of adhesion is uniform, the normal force converges towards the JKR solution (Fig.~\ref{fig:convergence_bem_rays}c)
This convergence behavior is consistent with reports in \citet{maugis_adhesion_1992}, \citet{greenwood_adhesion_1997} and \citet{muser_single-asperity_2014}.

For $n_\mathrm{rays} =4$ (Fig.~\ref{fig:convergence_bem_rays}a), the contact shapes of all crack-front models are similarly close to the BEM results.
While CF-K fits better in the peak and the valley of the undulation, CF-G and CF-E are closer to BEM between these
two locations.
For $n_\mathrm{rays}=64$ (Fig.~\ref{fig:convergence_bem_rays}b), the crack-front models differ more clearly
but the errors due to the finite interaction range in BEM are also relatively large.
However, as we mentioned in the previous paragraph,
the short-ranged limit has slightly higher contact radii than our reference simulation.
Therefore we conclude that for large $n_\mathrm{rays}$, the contact radius is significantly underestimated by CF-K.

The crack-front models unambiguously differ in the prediction of the normal force.
The CF-E model best captures how the normal force depends
on the heterogeneity wavelength (Fig.~\ref{fig:convergence_bem_rays}c)
and amplitude (Fig.~\ref{fig:convergence_bem_rays}d).
The deviations of CF-G from CF-E are small and disappear for small heterogeneities.
CL-K has the largest discrepancies and predicts less adhesive normal forces.
However, all errors are relatively small because the leading order contribution of $\Delta w$ to the normal forces
is of second order in BEM and in all the crack-front models.
This scaling has been pointed out by~\cite{argatov_controlling_2021} and is visible in Fig.~\ref{fig:convergence_bem_rays}d,
where we represent the normal force as a function of $\Delta w^2$.

We now discuss qualitatively how the crack shapes depend on $\Delta w$ (Fig.~\ref{fig:convergence_bem_rays}e).
In all crack-front models, the amplitude of the contact radius undulation is smaller than in BEM.
This is consistent with the numerical simulations of~\cite{fares_crack_1989}
and the second order perturbation by~\cite{leblond_second-order_2012}.
As in Fares' simulations, the relation between contact radius and work of adhesion (i.e. stress intensity factor) is
more nonlinear for the minimum contact radius than for the maximum contact radius.

\citet{leblond_second-order_2012} pointed out that for a semi-infinite crack ($n_\mathrm{rays} \to \infty$),
the mean contact radius depends only on the mean work of adhesion along the crack front.
Leblond's statement follows from the existence of an elastic potential and hence is satisfied by construction in CF-E.
In CF-G (Eq. \eqref{eq:G-lin}), the mean contact radius $\tilde a_0$
is independent of $\Delta w$ at any $n_\mathrm{rays}$;
in fact, CF-E and CF-G are identical in the limit $n_\mathrm{rays}\to \infty$.
CF-G conserves energy only in that limit.
In contrast, in the CF-K model (Eq.~\eqref{eq:K-lin}), $\tilde{a}_{0}$ decreases with increasing $\Delta w$.
The reason is that $\tilde a_0$ depends only on the average fracture toughness: taking the average of Eq.~\eqref{eq:K-lin}
yields $\left< K_\mathrm{c} \right> = K_\mathrm{J}(\tilde a_0)$.
At fixed $w_m$, $\left< K_\mathrm{c} \right> \propto \left< \sqrt{w} \right> $ decreases with increasing heterogeneity.
This contact radius offset is independent of $n_\mathrm{rays}$ and becomes large compared to the undulation when $n_\mathrm{rays}$ increases
(Fig.~\ref{fig:convergence_bem_rays}b).
Therefore, by predicting too small contact radii, CF-K violates energy conservation.
In summary,
for large $n_\mathrm{rays}$, the contact area is more accurate in CF-G and CF-E than in CF-K.
Because CF-K underestimates the contact area, it also predicts less adhesive normal forces.

The main benefit of CF-E is that it allows us to define rigorously the normal force by Eq.~\eqref{eq:normal_force_from_energy}.
Inserting the Taylor expansion of $F_\mathrm{J}$ into Eq.~\eqref{eq:circular_F},
only terms of second order in the amplitude of the contact radius remain\footnote{The first term in the Taylor expansion of $F_\mathrm{J}$ around $a(\theta) = \tilde a_0$ disappears by averaging over the perimeter.}.
$\partial^2 F_\mathrm{J}/\partial a^2$ and $\partial G_\text{J}/\partial \Delta$ are both negative, so that small-scale fluctuations of the
contact radius lead to a more negative (more adhesive) force.
Hence, fluctuations of the work of adhesion along the contact line slightly
increase adhesion, which is in agreement with the BEM results in Fig.~\ref{fig:convergence_bem_rays}d.
However, this effect vanishes for small heterogeneities (Fig.~\ref{fig:convergence_bem_rays}c): for large $n$, $\tilde a_n \propto \Delta w / n$,
so that $F(a) - F_\mathrm{J}(\tilde a_0) \propto \sum_n |n| |\tilde a_n|^2 \to 0$.

To conclude, while all the crack-perturbation methods are valid first order approximations, the ansatz possessing
an elastic energy (CF-E) yields the best overall results.
At large $n_\mathrm{rays}$ (small heterogeneities), the
CF-G model and the CF-E model are equivalent.
The CF-G equation is simpler and can be used in analytical theories for crack pinning \citep{demery_microstructural_2014}.
For an amplitude of the heterogeneity of $\Delta w \leq 0.4$, the errors
due to the finite interaction range in expensive BEM simulations are comparable to the errors of the first order perturbation models.

\section{Application to a random heterogeneity}

\begin{figure}
	\includegraphics[width=\textwidth]{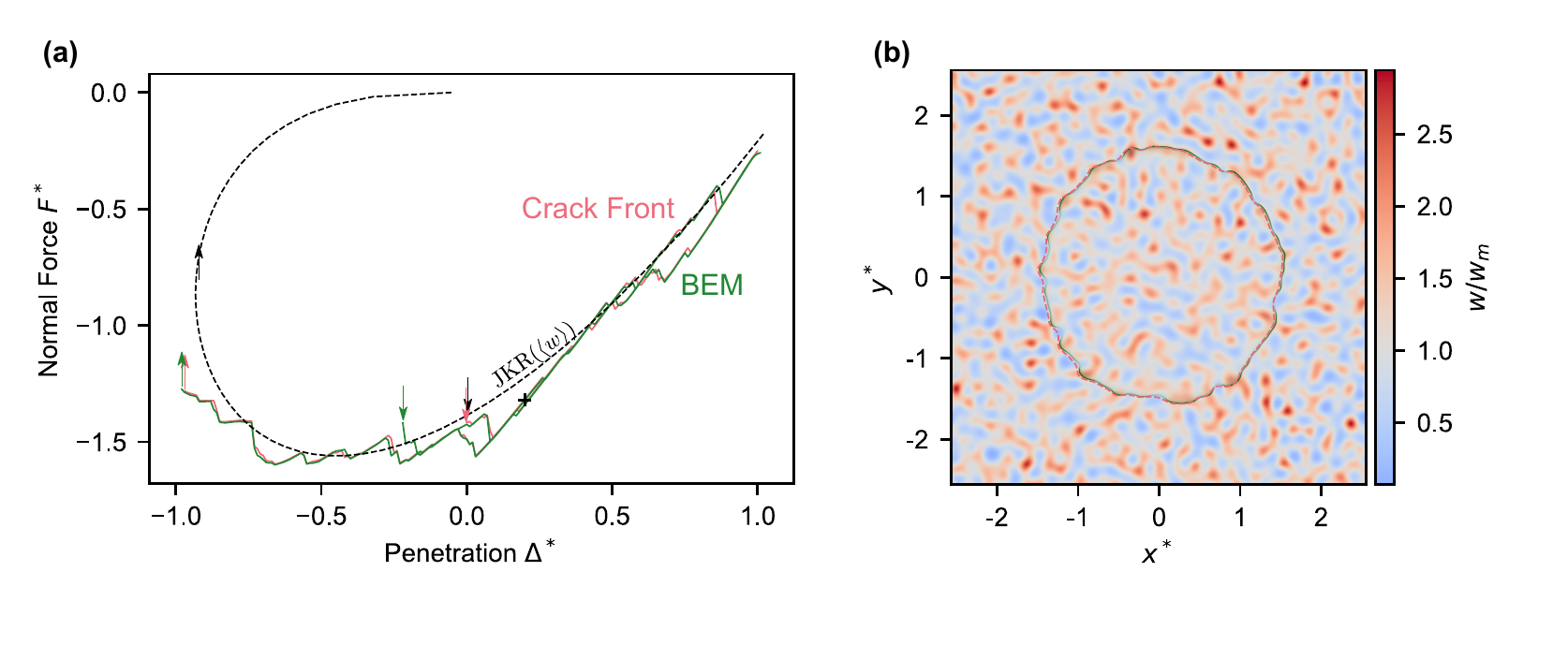}
	\caption{
		(a) Force-penetration curves from a boundary element method (BEM) and a crack-front simulation (CF-E)
		on the spatially random work of adhesion shown in panel (b).
		We also show the prediction by the JKR model for a homogeneous surface at the same average work of adhesion.
		The arrows indicate the jump into contact and the jump out of contact instabilities.
		(b) Contact areas at the penetration $\Delta^*=0.2$, indicated by the cross in the force-penetration curve,
		on top of the work of adhesion field.
		The tensile pressures of the contact mechanics simulation are shown in green,
		so that the perimeter of the contact is indicated by the darkest green pixels.
		The pink dashed line is the contact perimeter calculated with the crack-front model.
		The work of adhesion field has a standard deviation $\simeq 0.4 w_m$ and a short wavelength cutoff at $\ell_\mathrm{het}^* = 0.2$.
		We used the median work of adhesion $w_m = \left< K_\mathrm{c} \right>^2 / 2 E^\prime$ for nondimensionalization.
	}
	\label{fig:penforcewithinset_random}
\end{figure}

To illustrate the usefulness of the crack-front model to study the adhesion hysteresis, we simulated the indentation
and retraction of a sphere with randomly fluctuating work of adhesion.
We compare the results of BEM and the CF-E model in Fig.~\ref{fig:penforcewithinset_random}.
The crack-front simulation yields almost the same result as BEM, but requires only one minute instead of one day of computational time.
(Both simulations were performed on a single core.)

The BEM simulation was discretized on a $1024 \times 1024$ grid with pixel size $\ell_\mathrm{pix}^* = 0.005$.
On the same grid, we generated a random Gaussian toughness field with mean $\left< K_\mathrm{c}\right> = \sqrt{2 E^\prime w_m}$ and
standard deviation $K_{c, \text{rms}} = 0.2 \left< K_\mathrm{c} \right>$ (Fig.~\ref{fig:penforcewithinset_random}b).
We chose a Gaussian fracture toughness field instead of a work of adhesion field because it is easier to  avoid negative work of adhesion values.
Because the work of adhesion is the square of a Gaussian field,
the mean $\left< w \right>$ exceeds the median $w_m$ by a factor 1.04.
The standard deviation $w_\mathrm{rms} \simeq 0.39 w_m$.
The Fourier spectrum is flat at wavelengths above the correlation length $\ell_\mathrm{het}^*=0.2$ and 0 below.
The interaction range corresponds to a cohesive zone size $\ell^*_\mathrm{coz} \simeq 0.012$.

We performed a crack-front simulation on the same work of adhesion field.
In order to evaluate the equilibrium condition of the crack front (Eq.~\eqref{eq:CF-E_equilibrium}),
we interpolated the work of adhesion with bicubic splines between the grid points.
We used 512 collocation points on the crack front so that the largest spacing between the collocation points
(at the maximal penetration) is $\simeq \ell_\mathrm{het}^* / 10$.

In both simulations, we increased the penetration $\Delta^*$ in steps of 0.01
until the maximum penetration $\Delta^*_\mathrm{max} = 1$ was reached
and then decreased it until pull off.
The starting penetration was chosen to be lower than the jump into contact instability.

In Fig.~\ref{fig:penforcewithinset_random}a, the force-penetration curves computed with BEM and the crack-front model nearly overlap.
For reference, we also show the force-penetration curve resulting from a homogeneous work of adhesion (JKR model)
having the same mean value.
The homogeneous contact has one hysteresis loop corresponding to the jump into and the jump out of contact instabilities; the force is reversible at positive penetrations.
In our simulations, the heterogeneities are energy barriers that pin the crack front and alter the normal force.
The kinks in the force curve correspond to depinning instabilities that dissipate energy and lead to additional hysteresis loops \citep{joanny_model_1984}.
The adhesion hysteresis caused by crack-front pinning will be discussed in more detail in an upcoming publication.

We now discuss the differences between the crack front and BEM results.
In BEM, the jump into contact instability occurs too early: it converges much slower with interaction range than other quantities
\citep{wu_jump--contact_2010, ciavarella_effect_2017, wang_modeling_2021}. The remainder of the force-penetration curve,
including depinning instabilities, is well converged.
Other discrepancies in the force-penetration curves are due to the linearization in the crack-front model.
The contact perimeters agree well, with the most significant deviations in the regions with low work of adhesion.
This is consistent with what we observed in the previous section.
The two simulations deviate more significantly at a few penetration values, where the instabilities occur at slightly different penetrations.
An animation comparing the contact shape of BEM and the contact line during the whole indentation retraction process
is provided in the supplementary material.

As mentioned at the beginning of this section, the crack-front simulations are computationally much cheaper than the BEM simulations.
Furthermore, in BEM and in the crack-front model, the number of pixels in the linear dimension $n$ scales with $\ell_\mathrm{het}^{-1}$ (the total number of pixels in BEM
is then $\sim n^2$).
While in BEM, the computation time of the elastic deformations increases as $n^2 \log n$ (two dimensional fast Fourier transform),
in the crack-front model it only increases as $n \log n$ (one dimensional fast Fourier transform).
The crack-front model will enable us to simulate smaller heterogeneities and softer spheres than possible with traditional BEM approaches.

\section{Conclusion}

We described the adhesion of a sphere against a chemically heterogeneous surface by first-order crack perturbation and
validated this model against a boundary element method (BEM).
We compared different variants of the first-order perturbation and found that the best approach is to approximate the energy quadratically.
Linearizing the energy release rate is equivalent in the limit of small heterogeneity size, but linearizing the stress intensity factor underestimates adhesion in that limit.

By its efficiency, the crack-front model allows us to simulate orders of magnitude
larger systems than possible with BEMs.
This simplified model requires, however, that the work of adhesion heterogeneity is sufficiently small
for the contact area to be nearly circular.
Adapting the crack-front model to surface roughness will allow us to test whether crack-front pinning can explain the
role of surface roughness in adhesion hysteresis \citep{dalvi_linking_2019}
and to make theoretical predictions based on previous work
on pinning of elastic lines by a random field \citep{demery_microstructural_2014}.

\section*{Acknowledgments}

We thank W. Beck Andrews, Ali Dhinojwala, Patrick Dondl, Andreas Greiner and Tevis D. B. Jacobs for useful discussion.
We thank Sindhu Singh for implementing the constrained conjugate-gradients algorithm
and Laura Mahoney for writing assistance.
We are indebted to Mark O. Robbins for pointing us to the analogy to an elastic line.
Funding was provided by the Deutsche Forschungsgemeinschaft (DFG, German Research Foundation)
under Germany’s Excellence Strategy – EXC-2193/1 – 390951807 and by the European Research Council (StG-757343).
Numerical simulations were performed on bwForCluster NEMO (University of Freiburg, DFG grant INST 39/963-1 FUGG).


\begin{thebibliography}{105}
\expandafter\ifx\csname natexlab\endcsname\relax\def\natexlab#1{#1}\fi
\providecommand{\url}[1]{\texttt{#1}}
\providecommand{\href}[2]{#2}
\providecommand{\path}[1]{#1}
\providecommand{\DOIprefix}{doi:}
\providecommand{\ArXivprefix}{arXiv:}
\providecommand{\URLprefix}{URL: }
\providecommand{\Pubmedprefix}{pmid:}
\providecommand{\doi}[1]{\href{http://dx.doi.org/#1}{\path{#1}}}
\providecommand{\Pubmed}[1]{\href{pmid:#1}{\path{#1}}}
\providecommand{\bibinfo}[2]{#2}
\ifx\xfnm\relax \def\xfnm[#1]{\unskip,\space#1}\fi
\bibitem[{Amaral et~al.(1995)Amaral, Barabasi, Makse and
  Stanley}]{amaral_scaling_1995}
\bibinfo{author}{Amaral, L.A.N.}, \bibinfo{author}{Barabasi, A.L.},
  \bibinfo{author}{Makse, H.A.}, \bibinfo{author}{Stanley, H.E.},
  \bibinfo{year}{1995}.
\newblock \bibinfo{title}{Scaling properties of driven interfaces in disordered
  media}.
\newblock \bibinfo{journal}{Phys. Rev. E} \bibinfo{volume}{52},
  \bibinfo{pages}{4087}.
\newblock \DOIprefix\doi{10.1103/PhysRevE.52.4087}.
\bibitem[{Argatov(2021)}]{argatov_controlling_2021}
\bibinfo{author}{Argatov, I.I.}, \bibinfo{year}{2021}.
\newblock \bibinfo{title}{Controlling the adhesive pull-off force via the
  change of contact geometry}.
\newblock \bibinfo{journal}{Philos. Trans. R. Soc. London, Ser. A}
  \bibinfo{volume}{379}, \bibinfo{pages}{20200392}.
\newblock \DOIprefix\doi{10.1098/rsta.2020.0392}.
\bibitem[{Barthel(2008)}]{barthel_adhesive_2008}
\bibinfo{author}{Barthel, E.}, \bibinfo{year}{2008}.
\newblock \bibinfo{title}{Adhesive elastic contacts: {JKR} and more}.
\newblock \bibinfo{journal}{J. Phys. D: Appl. Phys.} \bibinfo{volume}{41},
  \bibinfo{pages}{163001}.
\newblock \DOIprefix\doi{10.1088/0022-3727/41/16/163001}.
\bibitem[{Batrouni et~al.(2002)Batrouni, Hansen and
  Schmittbuhl}]{batrouni_heterogeneous_2002}
\bibinfo{author}{Batrouni, G.G.}, \bibinfo{author}{Hansen, A.},
  \bibinfo{author}{Schmittbuhl, J.}, \bibinfo{year}{2002}.
\newblock \bibinfo{title}{Heterogeneous interfacial failure between two elastic
  blocks}.
\newblock \bibinfo{journal}{Phys. Rev. E} \bibinfo{volume}{65},
  \bibinfo{pages}{036126}.
\newblock \DOIprefix\doi{10.1103/PhysRevE.65.036126}.
\bibitem[{Bazrafshan et~al.(2017)Bazrafshan, {de Rooij}, Valefi and
  Schipper}]{bazrafshan_numerical_2017}
\bibinfo{author}{Bazrafshan, M.}, \bibinfo{author}{{de Rooij}, M.B.},
  \bibinfo{author}{Valefi, M.}, \bibinfo{author}{Schipper, D.J.},
  \bibinfo{year}{2017}.
\newblock \bibinfo{title}{Numerical method for the adhesive normal contact
  analysis based on a {{Dugdale}} approximation}.
\newblock \bibinfo{journal}{Tribol. Int.} \bibinfo{volume}{112},
  \bibinfo{pages}{117--128}.
\newblock \DOIprefix\doi{10.1016/j.triboint.2017.04.001}.
\bibitem[{Bonamy et~al.(2008)Bonamy, Santucci and
  Ponson}]{bonamy_crackling_2008}
\bibinfo{author}{Bonamy, D.}, \bibinfo{author}{Santucci, S.},
  \bibinfo{author}{Ponson, L.}, \bibinfo{year}{2008}.
\newblock \bibinfo{title}{Crackling dynamics in material failure as the
  signature of a self-organized dynamic phase transition}.
\newblock \bibinfo{journal}{Phys. Rev. Lett.} \bibinfo{volume}{101},
  \bibinfo{pages}{045501}.
\newblock \DOIprefix\doi{10.1103/PhysRevLett.101.045501}.
\bibitem[{Borodachev(1991)}]{borodachev_contact_1991}
\bibinfo{author}{Borodachev, N.M.}, \bibinfo{year}{1991}.
\newblock \bibinfo{title}{Contact problem for an elastic half-space with a
  near-circular contact area}.
\newblock \bibinfo{journal}{Soviet Appl. Mech.} \bibinfo{volume}{27},
  \bibinfo{pages}{118--123}.
\newblock \DOIprefix\doi{10.1007/BF00887799}.
\bibitem[{Briggs and Briscoe(1977)}]{briggs_effect_1977}
\bibinfo{author}{Briggs, G.A.D.}, \bibinfo{author}{Briscoe, B.J.},
  \bibinfo{year}{1977}.
\newblock \bibinfo{title}{The effect of surface topography on the adhesion of
  elastic solids}.
\newblock \bibinfo{journal}{J. Phys. D: Appl. Phys.} \bibinfo{volume}{10},
  \bibinfo{pages}{2453--2466}.
\newblock \DOIprefix\doi{10.1088/0022-3727/10/18/010}.
\bibitem[{Bugnicourt et~al.(2018)Bugnicourt, Sainsot, Dureisseix, Gauthier and
  Lubrecht}]{bugnicourt_fft-based_2018}
\bibinfo{author}{Bugnicourt, R.}, \bibinfo{author}{Sainsot, P.},
  \bibinfo{author}{Dureisseix, D.}, \bibinfo{author}{Gauthier, C.},
  \bibinfo{author}{Lubrecht, A.A.}, \bibinfo{year}{2018}.
\newblock \bibinfo{title}{{{FFT}}-based methods for solving a rough adhesive
  contact: Description and convergence study}.
\newblock \bibinfo{journal}{Tribol. Lett.} \bibinfo{volume}{66},
  \bibinfo{pages}{29}.
\newblock \DOIprefix\doi{10.1007/s11249-017-0980-z}.
\bibitem[{Byrd et~al.(1995)Byrd, Lu, Nocedal and Zhu}]{byrd_limited_1995}
\bibinfo{author}{Byrd, R.H.}, \bibinfo{author}{Lu, P.},
  \bibinfo{author}{Nocedal, J.}, \bibinfo{author}{Zhu, C.},
  \bibinfo{year}{1995}.
\newblock \bibinfo{title}{A limited memory algorithm for bound constrained
  optimization}.
\newblock \bibinfo{journal}{SIAM J. Sci. Comput.} \bibinfo{volume}{16},
  \bibinfo{pages}{1190--1208}.
\newblock \DOIprefix\doi{10.1137/0916069}.
\bibitem[{Campa{\~n}{\'a} and M{\"u}ser(2006)}]{campana_practical_2006}
\bibinfo{author}{Campa{\~n}{\'a}, C.}, \bibinfo{author}{M{\"u}ser, M.H.},
  \bibinfo{year}{2006}.
\newblock \bibinfo{title}{Practical {{Green}}'s function approach to the
  simulation of elastic semi-infinite solids}.
\newblock \bibinfo{journal}{Phys. Rev. B} \bibinfo{volume}{74},
  \bibinfo{pages}{075420}.
\newblock \DOIprefix\doi{10.1103/PhysRevB.74.075420}.
\bibitem[{Carbone et~al.(2015)Carbone, Pierro and
  Recchia}]{carbone_loading-unloading_2015}
\bibinfo{author}{Carbone, G.}, \bibinfo{author}{Pierro, E.},
  \bibinfo{author}{Recchia, G.}, \bibinfo{year}{2015}.
\newblock \bibinfo{title}{Loading-unloading hysteresis loop of randomly rough
  adhesive contacts}.
\newblock \bibinfo{journal}{Phys. Rev. E} \bibinfo{volume}{92},
  \bibinfo{pages}{062404}.
\newblock \DOIprefix\doi{10.1103/PhysRevE.92.062404}.
\bibitem[{Chen et~al.(2008)Chen, Shi and Gao}]{chen_apparent_2008}
\bibinfo{author}{Chen, B.}, \bibinfo{author}{Shi, X.}, \bibinfo{author}{Gao,
  H.}, \bibinfo{year}{2008}.
\newblock \bibinfo{title}{Apparent fracture/adhesion energy of interfaces with
  periodic cohesive interactions}.
\newblock \bibinfo{journal}{Proc. R. Soc. London, Ser. A}
  \bibinfo{volume}{464}, \bibinfo{pages}{657--671}.
\newblock \DOIprefix\doi{10.1098/rspa.2007.0240}.
\bibitem[{Chen et~al.(1991)Chen, Helm and Israelachvili}]{chen_molecular_1991}
\bibinfo{author}{Chen, Y.L.}, \bibinfo{author}{Helm, C.A.},
  \bibinfo{author}{Israelachvili, J.N.}, \bibinfo{year}{1991}.
\newblock \bibinfo{title}{Molecular mechanisms associated with adhesion and
  contact angle hysteresis of monolayer surfaces}.
\newblock \bibinfo{journal}{J. Phys. Chem.} \bibinfo{volume}{95},
  \bibinfo{pages}{10736--10747}.
\newblock \DOIprefix\doi{10.1021/j100179a041}.
\bibitem[{Chopin et~al.(2015)Chopin, Boudaoud and
  {Adda-Bedia}}]{chopin_morphology_2015}
\bibinfo{author}{Chopin, J.}, \bibinfo{author}{Boudaoud, A.},
  \bibinfo{author}{{Adda-Bedia}, M.}, \bibinfo{year}{2015}.
\newblock \bibinfo{title}{Morphology and dynamics of a crack front propagating
  in a model disordered material}.
\newblock \bibinfo{journal}{J. Mech. Phys. Solids} \bibinfo{volume}{74},
  \bibinfo{pages}{38--48}.
\newblock \DOIprefix\doi{10.1016/j.jmps.2014.10.001}.
\bibitem[{Ciavarella et~al.(2017)Ciavarella, Greenwood and
  Barber}]{ciavarella_effect_2017}
\bibinfo{author}{Ciavarella, M.}, \bibinfo{author}{Greenwood, J.A.},
  \bibinfo{author}{Barber, J.R.}, \bibinfo{year}{2017}.
\newblock \bibinfo{title}{Effect of {{Tabor}} parameter on hysteresis losses
  during adhesive contact}.
\newblock \bibinfo{journal}{J. Mech. Phys. Solids} \bibinfo{volume}{98},
  \bibinfo{pages}{236--244}.
\newblock \DOIprefix\doi{10.1016/j.jmps.2016.10.005}.
\bibitem[{Dalvi et~al.(2019)Dalvi, Gujrati, Khanal, Pastewka, Dhinojwala and
  Jacobs}]{dalvi_linking_2019}
\bibinfo{author}{Dalvi, S.}, \bibinfo{author}{Gujrati, A.},
  \bibinfo{author}{Khanal, S.R.}, \bibinfo{author}{Pastewka, L.},
  \bibinfo{author}{Dhinojwala, A.}, \bibinfo{author}{Jacobs, T.D.B.},
  \bibinfo{year}{2019}.
\newblock \bibinfo{title}{Linking energy loss in soft adhesion to surface
  roughness}.
\newblock \bibinfo{journal}{Proc. Natl. Acad. Sci. U.S.A.}
  \bibinfo{volume}{116}, \bibinfo{pages}{25484--25490}.
\newblock \DOIprefix\doi{10.1073/pnas.1913126116}.
\bibitem[{Dapp and M{\"u}ser(2015)}]{dapp_contact_2015}
\bibinfo{author}{Dapp, W.B.}, \bibinfo{author}{M{\"u}ser, M.H.},
  \bibinfo{year}{2015}.
\newblock \bibinfo{title}{Contact mechanics of and {{Reynolds}} flow through
  saddle points: {{On}} the coalescence of contact patches and the leakage rate
  through near-critical constrictions}.
\newblock \bibinfo{journal}{Europhys. Lett.} \bibinfo{volume}{109},
  \bibinfo{pages}{44001}.
\newblock \DOIprefix\doi{10.1209/0295-5075/109/44001}.
\bibitem[{D{\'e}mery et~al.(2014)D{\'e}mery, Rosso and
  Ponson}]{demery_microstructural_2014}
\bibinfo{author}{D{\'e}mery, V.}, \bibinfo{author}{Rosso, A.},
  \bibinfo{author}{Ponson, L.}, \bibinfo{year}{2014}.
\newblock \bibinfo{title}{From microstructural features to effective toughness
  in disordered brittle solids}.
\newblock \bibinfo{journal}{Europhys. Lett.} \bibinfo{volume}{105},
  \bibinfo{pages}{34003}.
\newblock \DOIprefix\doi{10.1209/0295-5075/105/34003}.
\bibitem[{Deng and Kesari(2017)}]{deng_molecular_2017}
\bibinfo{author}{Deng, W.}, \bibinfo{author}{Kesari, H.}, \bibinfo{year}{2017}.
\newblock \bibinfo{title}{Molecular statics study of depth-dependent hysteresis
  in nano-scale adhesive elastic contacts}.
\newblock \bibinfo{journal}{Modell. Simul. Mater. Sci. Eng.}
  \bibinfo{volume}{25}, \bibinfo{pages}{055002}.
\newblock \DOIprefix\doi{10.1088/1361-651X/aa6ef8}.
\bibitem[{Deng and Kesari(2019)}]{deng_depth-dependent_2019}
\bibinfo{author}{Deng, W.}, \bibinfo{author}{Kesari, H.}, \bibinfo{year}{2019}.
\newblock \bibinfo{title}{Depth-dependent hysteresis in adhesive elastic
  contacts at large surface roughness}.
\newblock \bibinfo{journal}{Sci. Rep.} \bibinfo{volume}{9},
  \bibinfo{pages}{1639}.
\newblock \DOIprefix\doi{10.1038/s41598-018-38212-z}.
\bibitem[{Fares(1989)}]{fares_crack_1989}
\bibinfo{author}{Fares, N.}, \bibinfo{year}{1989}.
\newblock \bibinfo{title}{Crack fronts trapped by arrays of obstacles:
  Numerical solutions based on surface integral representation}.
\newblock \bibinfo{journal}{J. Appl. Mech.} \bibinfo{volume}{56},
  \bibinfo{pages}{837--843}.
\newblock \DOIprefix\doi{10.1115/1.3176179}.
\bibitem[{Feng(2000)}]{feng_contact_2000}
\bibinfo{author}{Feng, J.Q.}, \bibinfo{year}{2000}.
\newblock \bibinfo{title}{Contact behavior of spherical elastic particles: A
  computational study of particle adhesion and deformations}.
\newblock \bibinfo{journal}{Colloids Surf., A} \bibinfo{volume}{172},
  \bibinfo{pages}{175--198}.
\newblock \DOIprefix\doi{10.1016/S0927-7757(00)00580-X}.
\bibitem[{Fisher(1983)}]{fisher_threshold_1983}
\bibinfo{author}{Fisher, D.S.}, \bibinfo{year}{1983}.
\newblock \bibinfo{title}{Threshold behavior of charge-density waves pinned by
  impurities}.
\newblock \bibinfo{journal}{Phys. Rev. Lett.} \bibinfo{volume}{50},
  \bibinfo{pages}{1486--1489}.
\newblock \DOIprefix\doi{10.1103/PhysRevLett.50.1486}.
\bibitem[{Fuller and Tabor(1975)}]{fuller_effect_1975}
\bibinfo{author}{Fuller, K.N.G.}, \bibinfo{author}{Tabor, D.},
  \bibinfo{year}{1975}.
\newblock \bibinfo{title}{The effect of surface roughness on the adhesion of
  elastic solids}.
\newblock \bibinfo{journal}{Proc. R. Soc. London, Ser. A}
  \bibinfo{volume}{345}, \bibinfo{pages}{327--342}.
\newblock \DOIprefix\doi{10.1098/rspa.1975.0138}.
\bibitem[{Gao and Rice(1987a)}]{gao_nearly_1987}
\bibinfo{author}{Gao, H.}, \bibinfo{author}{Rice, J.R.}, \bibinfo{year}{1987}a.
\newblock \bibinfo{title}{Nearly circular connections of elastic half spaces}.
\newblock \bibinfo{journal}{J. Appl. Mech.} \bibinfo{volume}{54},
  \bibinfo{pages}{627--634}.
\newblock \DOIprefix\doi{10.1115/1.3173080}.
\bibitem[{Gao and Rice(1987b)}]{gao_somewhat_1987}
\bibinfo{author}{Gao, H.}, \bibinfo{author}{Rice, J.R.}, \bibinfo{year}{1987}b.
\newblock \bibinfo{title}{Somewhat circular tensile cracks}.
\newblock \bibinfo{journal}{Int. Journal of Fract.} \bibinfo{volume}{33},
  \bibinfo{pages}{155--174}.
\newblock \DOIprefix\doi{10.1007/BF00013168}.
\bibitem[{Gao and Rice(1989)}]{gao_first-order_1989}
\bibinfo{author}{Gao, H.}, \bibinfo{author}{Rice, J.R.}, \bibinfo{year}{1989}.
\newblock \bibinfo{title}{A first-order perturbation analysis of crack trapping
  by arrays of obstacles}.
\newblock \bibinfo{journal}{J. Appl. Mech.} \bibinfo{volume}{56},
  \bibinfo{pages}{828--836}.
\newblock \DOIprefix\doi{10.1115/1.3176178}.
\bibitem[{Ghanbarzadeh et~al.(2020)Ghanbarzadeh, Faraji and
  Neville}]{ghanbarzadeh_deterministic_2020}
\bibinfo{author}{Ghanbarzadeh, A.}, \bibinfo{author}{Faraji, M.},
  \bibinfo{author}{Neville, A.}, \bibinfo{year}{2020}.
\newblock \bibinfo{title}{Deterministic normal contact of rough surfaces with
  adhesion using a surface integral method}.
\newblock \bibinfo{journal}{Proc. R. Soc. London, Ser. A}
  \bibinfo{volume}{476}, \bibinfo{pages}{20200281}.
\newblock \DOIprefix\doi{10.1098/rspa.2020.0281}.
\bibitem[{Gjerden et~al.(2013)Gjerden, Stormo and
  Hansen}]{gjerden_universality_2013}
\bibinfo{author}{Gjerden, K.S.}, \bibinfo{author}{Stormo, A.},
  \bibinfo{author}{Hansen, A.}, \bibinfo{year}{2013}.
\newblock \bibinfo{title}{Universality classes in constrained crack growth}.
\newblock \bibinfo{journal}{Phys. Rev. Lett.} \bibinfo{volume}{111},
  \bibinfo{pages}{135502}.
\newblock \DOIprefix\doi{10.1103/PhysRevLett.111.135502}.
\bibitem[{Gjerden et~al.(2014)Gjerden, Stormo and Hansen}]{gjerden_local_2014}
\bibinfo{author}{Gjerden, K.S.}, \bibinfo{author}{Stormo, A.},
  \bibinfo{author}{Hansen, A.}, \bibinfo{year}{2014}.
\newblock \bibinfo{title}{Local dynamics of a randomly pinned crack front: A
  numerical study}.
\newblock \bibinfo{journal}{Front. Phys.} \bibinfo{volume}{2},
  \bibinfo{pages}{66}.
\newblock \DOIprefix\doi{10.3389/fphy.2014.00066}.
\bibitem[{Gladwell(2008)}]{gladwell_contact_2008}
\bibinfo{editor}{Gladwell, G.M.L.} (Ed.), \bibinfo{year}{2008}.
\newblock \bibinfo{title}{Contact Problems}. volume \bibinfo{volume}{155} of
  \textit{\bibinfo{series}{Solid Mechanics and Its Applications}}.
\newblock \bibinfo{publisher}{{Springer Netherlands}},
  \bibinfo{address}{{Dordrecht}}.
\newblock \DOIprefix\doi{10.1007/978-1-4020-9043-1}.
\bibitem[{Greenwood(1997)}]{greenwood_adhesion_1997}
\bibinfo{author}{Greenwood, J.A.}, \bibinfo{year}{1997}.
\newblock \bibinfo{title}{Adhesion of elastic spheres}.
\newblock \bibinfo{journal}{Proc. R. Soc. London, Ser. A}
  \bibinfo{volume}{453}, \bibinfo{pages}{1277--1297}.
\newblock \DOIprefix\doi{10.1098/rspa.1997.0070}.
\bibitem[{Greenwood(2017)}]{greenwood_reflections_2017}
\bibinfo{author}{Greenwood, J.A.}, \bibinfo{year}{2017}.
\newblock \bibinfo{title}{Reflections on and extensions of the {{Fuller}} and
  {{Tabor}} theory of rough surface adhesion}.
\newblock \bibinfo{journal}{Tribol. Lett.} \bibinfo{volume}{65},
  \bibinfo{pages}{159}.
\newblock \DOIprefix\doi{10.1007/s11249-017-0938-1}.
\bibitem[{Greenwood and Williamson(1966)}]{greenwood_contact_1966}
\bibinfo{author}{Greenwood, J.A.}, \bibinfo{author}{Williamson, J.B.P.},
  \bibinfo{year}{1966}.
\newblock \bibinfo{title}{Contact of nominally flat surfaces}.
\newblock \bibinfo{journal}{Proc. R. Soc. London, Ser. A}
  \bibinfo{volume}{295}, \bibinfo{pages}{300--319}.
\newblock \DOIprefix\doi{10.1098/rspa.1966.0242}.
\bibitem[{Griffith and Taylor(1921)}]{griffith_vi._1921}
\bibinfo{author}{Griffith, A.A.}, \bibinfo{author}{Taylor, G.I.},
  \bibinfo{year}{1921}.
\newblock \bibinfo{title}{{{VI}}. {{The}} phenomena of rupture and flow in
  solids}.
\newblock \bibinfo{journal}{Philos. Trans. R. Soc. London, Ser. A}
  \bibinfo{volume}{221}, \bibinfo{pages}{163--198}.
\newblock \DOIprefix\doi{10.1098/rsta.1921.0006}.
\bibitem[{Guduru(2007)}]{guduru_detachment_2007-1}
\bibinfo{author}{Guduru, P.R.}, \bibinfo{year}{2007}.
\newblock \bibinfo{title}{Detachment of a rigid solid from an elastic wavy
  surface: {{Theory}}}.
\newblock \bibinfo{journal}{J. Mech. Phys. Solids} \bibinfo{volume}{55},
  \bibinfo{pages}{445--472}.
\newblock \DOIprefix\doi{10.1016/j.jmps.2006.09.004}.
\bibitem[{Guduru and Bull(2007)}]{guduru_detachment_2007}
\bibinfo{author}{Guduru, P.R.}, \bibinfo{author}{Bull, C.},
  \bibinfo{year}{2007}.
\newblock \bibinfo{title}{Detachment of a rigid solid from an elastic wavy
  surface: {{Experiments}}}.
\newblock \bibinfo{journal}{J. Mech. Phys. Solids} \bibinfo{volume}{55},
  \bibinfo{pages}{473--488}.
\newblock \DOIprefix\doi{10.1016/j.jmps.2006.09.007}.
\bibitem[{Hertz(1881)}]{hertz_ueber_1881}
\bibinfo{author}{Hertz, H.}, \bibinfo{year}{1881}.
\newblock \bibinfo{title}{Ueber die {Ber\"uhrung} fester elastischer
  {K\"orper}}.
\newblock \bibinfo{journal}{J. Reine Angew. Math.} \bibinfo{volume}{92},
  \bibinfo{pages}{156--171}.
\newblock \DOIprefix\doi{10.1515/crll.1882.92.156}.
\bibitem[{Hockney(1970)}]{hockney_potential_1970}
\bibinfo{author}{Hockney, R.W.}, \bibinfo{year}{1970}.
\newblock \bibinfo{title}{The potential calculation and some applications}, in:
  \bibinfo{editor}{Alder, B.A.}, \bibinfo{editor}{Fernbach, S.},
  \bibinfo{editor}{Rotenberg, M.} (Eds.), \bibinfo{booktitle}{Methods in
  Computational Physics, {{Vol}}. 9}. \bibinfo{publisher}{Academic Press},
  \bibinfo{address}{{New York}}, pp. \bibinfo{pages}{135--211}.
\bibitem[{Hulikal et~al.(2017)Hulikal, Bhattacharya and
  Lapusta}]{hulikal_relation_2017}
\bibinfo{author}{Hulikal, S.}, \bibinfo{author}{Bhattacharya, K.},
  \bibinfo{author}{Lapusta, N.}, \bibinfo{year}{2017}.
\newblock \bibinfo{title}{The relation between a microscopic threshold-force
  model and macroscopic models of adhesion}.
\newblock \bibinfo{journal}{Acta Mech. Sin.} \bibinfo{volume}{33},
  \bibinfo{pages}{508--515}.
\newblock \DOIprefix\doi{10.1007/s10409-016-0630-y}.
\bibitem[{Irwin(1957)}]{irwin_analysis_1957}
\bibinfo{author}{Irwin, G.R.}, \bibinfo{year}{1957}.
\newblock \bibinfo{title}{Analysis of stresses and strains near the end of a
  crack transversing a plate}.
\newblock \bibinfo{journal}{J. Appl. Mech.} \bibinfo{volume}{24},
  \bibinfo{pages}{361--364}.
\newblock \DOIprefix\doi{10.1115/1.4011547}.
\bibitem[{Joanny and {de Gennes}(1984)}]{joanny_model_1984}
\bibinfo{author}{Joanny, J.F.}, \bibinfo{author}{{de Gennes}, P.G.},
  \bibinfo{year}{1984}.
\newblock \bibinfo{title}{A model for contact angle hysteresis}.
\newblock \bibinfo{journal}{J. Chem. Phys.} \bibinfo{volume}{81},
  \bibinfo{pages}{552--562}.
\newblock \DOIprefix\doi{10.1063/1.447337}.
\bibitem[{Johnson(1985)}]{johnson_contact_1985}
\bibinfo{author}{Johnson, K.L.}, \bibinfo{year}{1985}.
\newblock \bibinfo{title}{Contact Mechanics}.
\newblock \bibinfo{publisher}{Cambridge University Press}.
\bibitem[{Johnson et~al.(1971)Johnson, Kendall and
  Roberts}]{johnson_surface_1971}
\bibinfo{author}{Johnson, K.L.}, \bibinfo{author}{Kendall, K.},
  \bibinfo{author}{Roberts, A.D.}, \bibinfo{year}{1971}.
\newblock \bibinfo{title}{Surface energy and the contact of elastic solids}.
\newblock \bibinfo{journal}{Proc. R. Soc. London, Ser. A}
  \bibinfo{volume}{324}, \bibinfo{pages}{301--313}.
\newblock \DOIprefix\doi{10.1098/rspa.1971.0141}.
\bibitem[{Kesari et~al.(2010)Kesari, Doll, Pruitt, Cai and
  Lew}]{kesari_role_2010}
\bibinfo{author}{Kesari, H.}, \bibinfo{author}{Doll, J.C.},
  \bibinfo{author}{Pruitt, B.L.}, \bibinfo{author}{Cai, W.},
  \bibinfo{author}{Lew, A.J.}, \bibinfo{year}{2010}.
\newblock \bibinfo{title}{Role of surface roughness in hysteresis during
  adhesive elastic contact}.
\newblock \bibinfo{journal}{Philos. Mag. Lett.} \bibinfo{volume}{90},
  \bibinfo{pages}{891--902}.
\newblock \DOIprefix\doi{10.1080/09500839.2010.521204}.
\bibitem[{Kesari and Lew(2011)}]{kesari_effective_2011}
\bibinfo{author}{Kesari, H.}, \bibinfo{author}{Lew, A.J.},
  \bibinfo{year}{2011}.
\newblock \bibinfo{title}{Effective macroscopic adhesive contact behavior
  induced by small surface roughness}.
\newblock \bibinfo{journal}{J. Mech. Phys. Solids} \bibinfo{volume}{59},
  \bibinfo{pages}{2488--2510}.
\newblock \DOIprefix\doi{10.1016/j.jmps.2011.07.009}.
\bibitem[{Larkin and Ovchinnikov(1979)}]{larkin_pinning_1979}
\bibinfo{author}{Larkin, A.I.}, \bibinfo{author}{Ovchinnikov, Y.N.},
  \bibinfo{year}{1979}.
\newblock \bibinfo{title}{Pinning in type {{II}} superconductors}.
\newblock \bibinfo{journal}{J. Low. Temp. Phys.} \bibinfo{volume}{34},
  \bibinfo{pages}{409--428}.
\newblock \DOIprefix\doi{10.1007/BF00117160}.
\bibitem[{Lazarus(2011)}]{lazarus_perturbation_2011}
\bibinfo{author}{Lazarus, V.}, \bibinfo{year}{2011}.
\newblock \bibinfo{title}{Perturbation approaches of a planar crack in linear
  elastic fracture mechanics: {{A}} review}.
\newblock \bibinfo{journal}{J. Mech. Phys. Solids} \bibinfo{volume}{59},
  \bibinfo{pages}{121--144}.
\newblock \DOIprefix\doi{10.1016/j.jmps.2010.12.006}.
\bibitem[{Lebihain et~al.(2021)Lebihain, Ponson, Kondo and
  Leblond}]{lebihain_effective_2021}
\bibinfo{author}{Lebihain, M.}, \bibinfo{author}{Ponson, L.},
  \bibinfo{author}{Kondo, D.}, \bibinfo{author}{Leblond, J.B.},
  \bibinfo{year}{2021}.
\newblock \bibinfo{title}{Effective toughness of disordered brittle solids:
  {{A}} homogenization framework}.
\newblock \bibinfo{journal}{J. Mech. Phys. Solids} \bibinfo{volume}{153},
  \bibinfo{pages}{104463}.
\newblock \DOIprefix\doi{10.1016/j.jmps.2021.104463}.
\bibitem[{Leblond et~al.(2012)Leblond, Patinet, Frelat and
  Lazarus}]{leblond_second-order_2012}
\bibinfo{author}{Leblond, J.B.}, \bibinfo{author}{Patinet, S.},
  \bibinfo{author}{Frelat, J.}, \bibinfo{author}{Lazarus, V.},
  \bibinfo{year}{2012}.
\newblock \bibinfo{title}{Second-order coplanar perturbation of a semi-infinite
  crack in an infinite body}.
\newblock \bibinfo{journal}{Eng. Fract. Mech.} \bibinfo{volume}{90},
  \bibinfo{pages}{129--142}.
\newblock \DOIprefix\doi{10.1016/j.engfracmech.2012.03.002}.
\bibitem[{Li et~al.(2019)Li, Pohrt and Popov}]{li_adhesive_2019}
\bibinfo{author}{Li, Q.}, \bibinfo{author}{Pohrt, R.}, \bibinfo{author}{Popov,
  V.L.}, \bibinfo{year}{2019}.
\newblock \bibinfo{title}{Adhesive strength of contacts of rough spheres}.
\newblock \bibinfo{journal}{Front. Mech. Eng.} \bibinfo{volume}{5},
  \bibinfo{pages}{7}.
\newblock \DOIprefix\doi{10.3389/fmech.2019.00007}.
\bibitem[{Liu et~al.(2000)Liu, Wang and Liu}]{liu_versatile_2000}
\bibinfo{author}{Liu, S.}, \bibinfo{author}{Wang, Q.}, \bibinfo{author}{Liu,
  G.}, \bibinfo{year}{2000}.
\newblock \bibinfo{title}{A versatile method of discrete convolution and
  {{FFT}} ({{DC}}-{{FFT}}) for contact analyses}.
\newblock \bibinfo{journal}{Wear} \bibinfo{volume}{243},
  \bibinfo{pages}{101--111}.
\newblock \DOIprefix\doi{10.1016/S0043-1648(00)00427-0}.
\bibitem[{Love(1929)}]{love_stress_1929}
\bibinfo{author}{Love, A.E.H.}, \bibinfo{year}{1929}.
\newblock \bibinfo{title}{{{IX}}. {{The}} stress produced in a semi-infinite
  solid by pressure on part of the boundary}.
\newblock \bibinfo{journal}{Philos. Trans. R. Soc. Lond. A}
  \bibinfo{volume}{228}, \bibinfo{pages}{377--420}.
\newblock \DOIprefix\doi{10.1098/rsta.1929.0009}.
\bibitem[{Maugis(1992)}]{maugis_adhesion_1992}
\bibinfo{author}{Maugis, D.}, \bibinfo{year}{1992}.
\newblock \bibinfo{title}{Adhesion of spheres: {{The JKR}}-{{DMT}} transition
  using a {{Dugdale}} model}.
\newblock \bibinfo{journal}{J. Colloid Interface Sci.} \bibinfo{volume}{150},
  \bibinfo{pages}{243--269}.
\newblock \DOIprefix\doi{10.1016/0021-9797(92)90285-T}.
\bibitem[{Maugis(2010)}]{maugis_contact_2010}
\bibinfo{author}{Maugis, D.}, \bibinfo{year}{2010}.
\newblock \bibinfo{title}{Contact, Adhesion and Rupture of Elastic Solids}.
\newblock \bibinfo{publisher}{{Springer}}, \bibinfo{address}{{Berlin; New
  York}}.
\newblock \DOIprefix\doi{10.1007/978-3-662-04125-3}.
\bibitem[{Medina and Dini(2014)}]{medina_numerical_2014}
\bibinfo{author}{Medina, S.}, \bibinfo{author}{Dini, D.}, \bibinfo{year}{2014}.
\newblock \bibinfo{title}{A numerical model for the deterministic analysis of
  adhesive rough contacts down to the nano-scale}.
\newblock \bibinfo{journal}{Int. J. Solids Struct.} \bibinfo{volume}{51},
  \bibinfo{pages}{2620--2632}.
\newblock \DOIprefix\doi{10.1016/j.ijsolstr.2014.03.033}.
\bibitem[{Middleton(1992)}]{middleton_asymptotic_1992}
\bibinfo{author}{Middleton, A.A.}, \bibinfo{year}{1992}.
\newblock \bibinfo{title}{Asymptotic uniqueness of the sliding state for
  charge-density waves}.
\newblock \bibinfo{journal}{Phys. Rev. Lett.} \bibinfo{volume}{68},
  \bibinfo{pages}{670--673}.
\newblock \DOIprefix\doi{10.1103/PhysRevLett.68.670}.
\bibitem[{Monti et~al.(2019)Monti, McGuiggan and Robbins}]{monti_effect_2019}
\bibinfo{author}{Monti, J.}, \bibinfo{author}{McGuiggan, P.M.},
  \bibinfo{author}{Robbins, M.O.}, \bibinfo{year}{2019}.
\newblock \bibinfo{title}{Effect of roughness and elasticity on interactions
  between charged colloidal spheres}.
\newblock \bibinfo{journal}{Langmuir} \bibinfo{volume}{35},
  \bibinfo{pages}{15948--15959}.
\newblock \DOIprefix\doi{10.1021/acs.langmuir.9b02161}.
\bibitem[{Monti et~al.(2021)Monti, Sanner and
  Pastewka}]{monti_distribution_2021}
\bibinfo{author}{Monti, J.M.}, \bibinfo{author}{Sanner, A.},
  \bibinfo{author}{Pastewka, L.}, \bibinfo{year}{2021}.
\newblock \bibinfo{title}{Distribution of gaps and adhesive interaction between
  contacting rough surfaces}.
\newblock \bibinfo{journal}{Tribol. Lett.} \bibinfo{volume}{69},
  \bibinfo{pages}{80}.
\newblock \DOIprefix\doi{10.1007/s11249-021-01454-6}.
\bibitem[{Muller et~al.(1980)Muller, Yushchenko and
  Derjaguin}]{muller_influence_1980}
\bibinfo{author}{Muller, V.M.}, \bibinfo{author}{Yushchenko, V.S.},
  \bibinfo{author}{Derjaguin, B.V.}, \bibinfo{year}{1980}.
\newblock \bibinfo{title}{On the influence of molecular forces on the
  deformation of an elastic sphere and its sticking to a rigid plane}.
\newblock \bibinfo{journal}{J. Colloid Interface Sci.} \bibinfo{volume}{77},
  \bibinfo{pages}{91--101}.
\newblock \DOIprefix\doi{10.1016/0021-9797(80)90419-1}.
\bibitem[{M{\"u}ser(2014)}]{muser_single-asperity_2014}
\bibinfo{author}{M{\"u}ser, M.H.}, \bibinfo{year}{2014}.
\newblock \bibinfo{title}{Single-asperity contact mechanics with positive and
  negative work of adhesion: {{Influence}} of finite-range interactions and a
  continuum description for the squeeze-out of wetting fluids}.
\newblock \bibinfo{journal}{Beilstein J. Nanotechnol.} \bibinfo{volume}{5},
  \bibinfo{pages}{419--437}.
\newblock \DOIprefix\doi{10.3762/bjnano.5.50}.
\bibitem[{M{\"u}ser(2016)}]{muser_dimensionless_2016}
\bibinfo{author}{M{\"u}ser, M.H.}, \bibinfo{year}{2016}.
\newblock \bibinfo{title}{A dimensionless measure for adhesion and effects of
  the range of adhesion in contacts of nominally flat surfaces}.
\newblock \bibinfo{journal}{Tribol. Int.} \bibinfo{volume}{100},
  \bibinfo{pages}{41--47}.
\newblock \DOIprefix\doi{10.1016/j.triboint.2015.11.010}.
\bibitem[{M{\"u}ser et~al.(2017)M{\"u}ser, Dapp, Bugnicourt, Sainsot, Lesaffre,
  Lubrecht, Persson, Harris, Bennett, Schulze, Rohde, Ifju, Gregory~Sawyer,
  Angelini, Esfahani, Kadkhodaei, Akbarzadeh, Wu, Vorlaufer, Vernes, Solhjoo,
  Vakis, Jackson, Xu, Streator, Rostami, Dini, Medina, Carbone, Bottiglione,
  Afferrante, Monti, Pastewka, Robbins and Greenwood}]{muser_meeting_2017}
\bibinfo{author}{M{\"u}ser, M.H.}, \bibinfo{author}{Dapp, W.B.},
  \bibinfo{author}{Bugnicourt, R.}, \bibinfo{author}{Sainsot, P.},
  \bibinfo{author}{Lesaffre, N.}, \bibinfo{author}{Lubrecht, T.A.},
  \bibinfo{author}{Persson, B.N.J.}, \bibinfo{author}{Harris, K.},
  \bibinfo{author}{Bennett, A.}, \bibinfo{author}{Schulze, K.},
  \bibinfo{author}{Rohde, S.}, \bibinfo{author}{Ifju, P.},
  \bibinfo{author}{Gregory~Sawyer, W.}, \bibinfo{author}{Angelini, T.},
  \bibinfo{author}{Esfahani, H.A.}, \bibinfo{author}{Kadkhodaei, M.},
  \bibinfo{author}{Akbarzadeh, S.}, \bibinfo{author}{Wu, J.J.},
  \bibinfo{author}{Vorlaufer, G.}, \bibinfo{author}{Vernes, A.},
  \bibinfo{author}{Solhjoo, S.}, \bibinfo{author}{Vakis, A.I.},
  \bibinfo{author}{Jackson, R.L.}, \bibinfo{author}{Xu, Y.},
  \bibinfo{author}{Streator, J.}, \bibinfo{author}{Rostami, A.},
  \bibinfo{author}{Dini, D.}, \bibinfo{author}{Medina, S.},
  \bibinfo{author}{Carbone, G.}, \bibinfo{author}{Bottiglione, F.},
  \bibinfo{author}{Afferrante, L.}, \bibinfo{author}{Monti, J.},
  \bibinfo{author}{Pastewka, L.}, \bibinfo{author}{Robbins, M.O.},
  \bibinfo{author}{Greenwood, J.A.}, \bibinfo{year}{2017}.
\newblock \bibinfo{title}{Meeting the contact-mechanics challenge}.
\newblock \bibinfo{journal}{Tribol. Lett.} \bibinfo{volume}{65},
  \bibinfo{pages}{118}.
\newblock \DOIprefix\doi{10.1007/s11249-017-0900-2}.
\bibitem[{Nocedal and Wright(2006)}]{nocedal_numerical_2006}
\bibinfo{author}{Nocedal, J.}, \bibinfo{author}{Wright, S.J.},
  \bibinfo{year}{2006}.
\newblock \bibinfo{title}{Numerical Optimization}.
\newblock Springer Series in Operations Research. \bibinfo{edition}{2nd} ed.,
  \bibinfo{publisher}{{Springer}}, \bibinfo{address}{{New York}}.
\bibitem[{Pastewka and Robbins(2014)}]{pastewka_contact_2014}
\bibinfo{author}{Pastewka, L.}, \bibinfo{author}{Robbins, M.O.},
  \bibinfo{year}{2014}.
\newblock \bibinfo{title}{Contact between rough surfaces and a criterion for
  macroscopic adhesion}.
\newblock \bibinfo{journal}{Proc. Natl. Acad. Sci. U.S.A.}
  \bibinfo{volume}{111}, \bibinfo{pages}{3298--3303}.
\newblock \DOIprefix\doi{10.1073/pnas.1320846111}.
\bibitem[{Pastewka and Robbins(2016)}]{pastewka_contact_2016}
\bibinfo{author}{Pastewka, L.}, \bibinfo{author}{Robbins, M.O.},
  \bibinfo{year}{2016}.
\newblock \bibinfo{title}{Contact area of rough spheres: Large scale
  simulations and simple scaling laws}.
\newblock \bibinfo{journal}{Appl. Phys. Lett.} \bibinfo{volume}{108},
  \bibinfo{pages}{221601}.
\newblock \DOIprefix\doi{10.1063/1.4950802}.
\bibitem[{Pastewka et~al.(2012)Pastewka, Sharp and
  Robbins}]{pastewka_seamless_2012}
\bibinfo{author}{Pastewka, L.}, \bibinfo{author}{Sharp, T.A.},
  \bibinfo{author}{Robbins, M.O.}, \bibinfo{year}{2012}.
\newblock \bibinfo{title}{Seamless elastic boundaries for atomistic
  calculations}.
\newblock \bibinfo{journal}{Phys. Rev. B} \bibinfo{volume}{86},
  \bibinfo{pages}{075459}.
\newblock \DOIprefix\doi{10.1103/PhysRevB.86.075459}.
\bibitem[{Patinet et~al.(2013)Patinet, Alzate, Barthel, Dalmas, Vandembroucq
  and Lazarus}]{patinet_finite_2013}
\bibinfo{author}{Patinet, S.}, \bibinfo{author}{Alzate, L.},
  \bibinfo{author}{Barthel, E.}, \bibinfo{author}{Dalmas, D.},
  \bibinfo{author}{Vandembroucq, D.}, \bibinfo{author}{Lazarus, V.},
  \bibinfo{year}{2013}.
\newblock \bibinfo{title}{Finite size effects on crack front pinning at
  heterogeneous planar interfaces: Experimental, finite elements and
  perturbation approaches}.
\newblock \bibinfo{journal}{J. Mech. Phys. Solids} \bibinfo{volume}{61},
  \bibinfo{pages}{311--324}.
\newblock \DOIprefix\doi{10.1016/j.jmps.2012.10.012}.
\bibitem[{Persson(2002a)}]{persson_adhesion_2002-1}
\bibinfo{author}{Persson, B.N.J.}, \bibinfo{year}{2002}a.
\newblock \bibinfo{title}{Adhesion between an elastic body and a randomly rough
  hard surface}.
\newblock \bibinfo{journal}{Eur. Phys. J. E} \bibinfo{volume}{8},
  \bibinfo{pages}{385--401}.
\newblock \DOIprefix\doi{10.1140/epje/i2002-10025-1}.
\bibitem[{Persson(2002b)}]{persson_adhesion_2002}
\bibinfo{author}{Persson, B.N.J.}, \bibinfo{year}{2002}b.
\newblock \bibinfo{title}{Adhesion between elastic bodies with randomly rough
  surfaces}.
\newblock \bibinfo{journal}{Phys. Rev. Lett.} \bibinfo{volume}{89},
  \bibinfo{pages}{245502}.
\newblock \DOIprefix\doi{10.1103/PhysRevLett.89.245502}.
\bibitem[{Persson and Scaraggi(2014)}]{persson_theory_2014}
\bibinfo{author}{Persson, B.N.J.}, \bibinfo{author}{Scaraggi, M.},
  \bibinfo{year}{2014}.
\newblock \bibinfo{title}{Theory of adhesion: Role of surface roughness}.
\newblock \bibinfo{journal}{J. Chem. Phys.} \bibinfo{volume}{141},
  \bibinfo{pages}{124701}.
\newblock \DOIprefix\doi{10.1063/1.4895789}.
\bibitem[{Persson and Tosatti(2001)}]{persson_effect_2001}
\bibinfo{author}{Persson, B.N.J.}, \bibinfo{author}{Tosatti, E.},
  \bibinfo{year}{2001}.
\newblock \bibinfo{title}{The effect of surface roughness on the adhesion of
  elastic solids}.
\newblock \bibinfo{journal}{J. Chem. Phys.} \bibinfo{volume}{115},
  \bibinfo{pages}{5597--5610}.
\newblock \DOIprefix\doi{10.1063/1.1398300}.
\bibitem[{Pohrt and Popov(2015)}]{pohrt_adhesive_2015}
\bibinfo{author}{Pohrt, R.}, \bibinfo{author}{Popov, V.L.},
  \bibinfo{year}{2015}.
\newblock \bibinfo{title}{Adhesive contact simulation of elastic solids using
  local mesh-dependent detachment criterion in boundary elements method}.
\newblock \bibinfo{journal}{Facta Universitatis} \bibinfo{volume}{13},
  \bibinfo{pages}{3--10}.
\bibitem[{Polonsky and Keer(1999)}]{polonsky_numerical_1999}
\bibinfo{author}{Polonsky, I.A.}, \bibinfo{author}{Keer, L.M.},
  \bibinfo{year}{1999}.
\newblock \bibinfo{title}{A numerical method for solving rough contact problems
  based on the multi-level multi-summation and conjugate gradient techniques}.
\newblock \bibinfo{journal}{Wear} \bibinfo{volume}{231},
  \bibinfo{pages}{206--219}.
\newblock \DOIprefix\doi{10.1016/S0043-1648(99)00113-1}.
\bibitem[{Ponson(2016)}]{ponson_statistical_2016}
\bibinfo{author}{Ponson, L.}, \bibinfo{year}{2016}.
\newblock \bibinfo{title}{Statistical aspects in crack growth phenomena: How
  the fluctuations reveal the failure mechanisms}.
\newblock \bibinfo{journal}{Int. J. of Fract.} \bibinfo{volume}{201},
  \bibinfo{pages}{11--27}.
\newblock \DOIprefix\doi{10.1007/s10704-016-0117-7}.
\bibitem[{Ponson and Bonamy(2010)}]{ponson_crack_2010}
\bibinfo{author}{Ponson, L.}, \bibinfo{author}{Bonamy, D.},
  \bibinfo{year}{2010}.
\newblock \bibinfo{title}{Crack propagation in brittle heterogeneous solids:
  {{Material}} disorder and crack dynamics}.
\newblock \bibinfo{journal}{Int. J. Fract.} \bibinfo{volume}{162},
  \bibinfo{pages}{21--31}.
\newblock \DOIprefix\doi{10.1007/s10704-010-9481-x}.
\bibitem[{Rey et~al.(2017)Rey, Anciaux and Molinari}]{rey_normal_2017}
\bibinfo{author}{Rey, V.}, \bibinfo{author}{Anciaux, G.},
  \bibinfo{author}{Molinari, J.F.}, \bibinfo{year}{2017}.
\newblock \bibinfo{title}{Normal adhesive contact on rough surfaces: Efficient
  algorithm for {{FFT}}-based {{BEM}} resolution}.
\newblock \bibinfo{journal}{Comput. Mech.} \bibinfo{volume}{60},
  \bibinfo{pages}{69--81}.
\newblock \DOIprefix\doi{10.1007/s00466-017-1392-5}.
\bibitem[{Rice(1989)}]{wei_weight_1989}
\bibinfo{author}{Rice, J.}, \bibinfo{year}{1989}.
\newblock \bibinfo{title}{Weight function theory for three-dimensional elastic
  crack analysis}, in: \bibinfo{editor}{Wei, R.}, \bibinfo{editor}{Gangloff,
  R.} (Eds.), \bibinfo{booktitle}{Fracture Mechanics: Perspectives and
  Directions (Twentieth Symposium)}. \bibinfo{publisher}{{{ASTM}}
  International}, \bibinfo{address}{{100 Barr Harbor Drive, PO Box C700, West
  Conshohocken, PA 19428-2959}}, pp. \bibinfo{pages}{29--57}.
\newblock \DOIprefix\doi{10.1520/STP18819S}.
\bibitem[{Rice(1985a)}]{rice_first-order_1985}
\bibinfo{author}{Rice, J.R.}, \bibinfo{year}{1985}a.
\newblock \bibinfo{title}{First-order variation in elastic fields due to
  variation in location of a planar crack front}.
\newblock \bibinfo{journal}{J. Appl. Mech.} \bibinfo{volume}{52},
  \bibinfo{pages}{571--579}.
\newblock \DOIprefix\doi{10.1115/1.3169103}.
\bibitem[{Rice(1985b)}]{rice_three-dimensional_1985}
\bibinfo{author}{Rice, J.R.}, \bibinfo{year}{1985}b.
\newblock \bibinfo{title}{Three-dimensional elastic crack tip interactions with
  transformation strains and dislocations}.
\newblock \bibinfo{journal}{Int. J. Solids Struct.} \bibinfo{volume}{21},
  \bibinfo{pages}{781--791}.
\newblock \DOIprefix\doi{10.1016/0020-7683(85)90081-2}.
\bibitem[{Robbins and Joanny(1987)}]{robbins_contact_1987}
\bibinfo{author}{Robbins, M.O.}, \bibinfo{author}{Joanny, J.F.},
  \bibinfo{year}{1987}.
\newblock \bibinfo{title}{Contact angle hysteresis on random surfaces}.
\newblock \bibinfo{journal}{Europhys. Lett.} \bibinfo{volume}{3},
  \bibinfo{pages}{729--735}.
\newblock \DOIprefix\doi{10.1209/0295-5075/3/6/013}.
\bibitem[{Rosso and Krauth(2002)}]{rosso_roughness_2002}
\bibinfo{author}{Rosso, A.}, \bibinfo{author}{Krauth, W.},
  \bibinfo{year}{2002}.
\newblock \bibinfo{title}{Roughness at the depinning threshold for a long-range
  elastic string}.
\newblock \bibinfo{journal}{Phys. Rev. E} \bibinfo{volume}{65},
  \bibinfo{pages}{025101}.
\newblock \DOIprefix\doi{10.1103/PhysRevE.65.025101}.
\bibitem[{Rosso et~al.(2007)Rosso, Le~Doussal and Wiese}]{rosso_numerical_2007}
\bibinfo{author}{Rosso, A.}, \bibinfo{author}{Le~Doussal, P.},
  \bibinfo{author}{Wiese, K.J.}, \bibinfo{year}{2007}.
\newblock \bibinfo{title}{Numerical calculation of the functional
  renormalization group fixed-point functions at the depinning transition}.
\newblock \bibinfo{journal}{Phys. Rev. B} \bibinfo{volume}{75},
  \bibinfo{pages}{220201}.
\newblock \DOIprefix\doi{10.1103/PhysRevB.75.220201}.
\bibitem[{Salvadori and Fantoni(2014)}]{salvadori_weight_2014}
\bibinfo{author}{Salvadori, A.}, \bibinfo{author}{Fantoni, F.},
  \bibinfo{year}{2014}.
\newblock \bibinfo{title}{Weight function theory and variational formulations
  for three-dimensional plane elastic cracks advancing}.
\newblock \bibinfo{journal}{Int. J. Solids Struct.} \bibinfo{volume}{51},
  \bibinfo{pages}{1030--1045}.
\newblock \DOIprefix\doi{10.1016/j.ijsolstr.2013.11.029}.
\bibitem[{Schmittbuhl et~al.(2003)Schmittbuhl, Hansen and
  Batrouni}]{schmittbuhl_roughness_2003}
\bibinfo{author}{Schmittbuhl, J.}, \bibinfo{author}{Hansen, A.},
  \bibinfo{author}{Batrouni, G.G.}, \bibinfo{year}{2003}.
\newblock \bibinfo{title}{Roughness of interfacial crack fronts:
  Stress-weighted percolation in the damage zone}.
\newblock \bibinfo{journal}{Phys. Rev. Lett.} \bibinfo{volume}{90},
  \bibinfo{pages}{045505}.
\newblock \DOIprefix\doi{10.1103/PhysRevLett.90.045505}.
\bibitem[{Schmittbuhl et~al.(1995)Schmittbuhl, Roux, Vilotte and {Jorgen M\aa
  loy K}}]{schmittbuhl_interfacial_1995}
\bibinfo{author}{Schmittbuhl, J.}, \bibinfo{author}{Roux, S.},
  \bibinfo{author}{Vilotte, J.P.}, \bibinfo{author}{{Jorgen M\aa loy K}},
  \bibinfo{year}{1995}.
\newblock \bibinfo{title}{Interfacial crack pinning: Effect of nonlocal
  interactions}.
\newblock \bibinfo{journal}{Phys. Rev. Lett.} \bibinfo{volume}{74},
  \bibinfo{pages}{1787--1790}.
\newblock \DOIprefix\doi{10.1103/PhysRevLett.74.1787}.
\bibitem[{Sneddon(1946)}]{sneddon_boussinesqs_1946}
\bibinfo{author}{Sneddon, I.N.}, \bibinfo{year}{1946}.
\newblock \bibinfo{title}{Boussinesq's problem for a flat-ended cylinder}.
\newblock \bibinfo{journal}{Math. Proc. Cambridge Philos. Soc.}
  \bibinfo{volume}{42}, \bibinfo{pages}{29--39}.
\newblock \DOIprefix\doi{10.1017/S0305004100022702}.
\bibitem[{Stanley and Kato(1997)}]{stanley_fft-based_1997}
\bibinfo{author}{Stanley, H.M.}, \bibinfo{author}{Kato, T.},
  \bibinfo{year}{1997}.
\newblock \bibinfo{title}{An {{FFT}}-based method for rough surface contact}.
\newblock \bibinfo{journal}{J. Tribol.} \bibinfo{volume}{119},
  \bibinfo{pages}{481--485}.
\newblock \DOIprefix\doi{10.1115/1.2833523}.
\bibitem[{Steihaug(1983)}]{steihaug_conjugate_1983}
\bibinfo{author}{Steihaug, T.}, \bibinfo{year}{1983}.
\newblock \bibinfo{title}{The conjugate gradient method and trust regions in
  large scale optimization}.
\newblock \bibinfo{journal}{SIAM J. Numer. Anal.} \bibinfo{volume}{20},
  \bibinfo{pages}{626--637}.
\newblock \DOIprefix\doi{10.1137/0720042}.
\bibitem[{Stormo et~al.(2012)Stormo, Gjerden and Hansen}]{stormo_onset_2012}
\bibinfo{author}{Stormo, A.}, \bibinfo{author}{Gjerden, K.S.},
  \bibinfo{author}{Hansen, A.}, \bibinfo{year}{2012}.
\newblock \bibinfo{title}{Onset of localization in heterogeneous interfacial
  failure}.
\newblock \bibinfo{journal}{Phys. Rev. E} \bibinfo{volume}{86},
  \bibinfo{pages}{025101}.
\newblock \DOIprefix\doi{10.1103/PhysRevE.86.025101}.
\bibitem[{Tabor(1977)}]{tabor_surface_1977}
\bibinfo{author}{Tabor, D.}, \bibinfo{year}{1977}.
\newblock \bibinfo{title}{Surface forces and surface interactions}.
\newblock \bibinfo{journal}{J. Colloid Interface Sci.} \bibinfo{volume}{58},
  \bibinfo{pages}{2--13}.
\newblock \DOIprefix\doi{10.1016/0021-9797(77)90366-6}.
\bibitem[{Tada et~al.(2000)Tada, Paris and Irwin}]{tada_stress_2000}
\bibinfo{author}{Tada, H.}, \bibinfo{author}{Paris, P.C.},
  \bibinfo{author}{Irwin, G.R.}, \bibinfo{year}{2000}.
\newblock \bibinfo{title}{The Stress Analysis Of Cracks Handbook}.
\newblock \bibinfo{edition}{3rd} ed., \bibinfo{publisher}{{{ASME}} Press},
  \bibinfo{address}{{New York}}.
\bibitem[{Tanguy and Vettorel(2004)}]{tanguy_weak_2004}
\bibinfo{author}{Tanguy, A.}, \bibinfo{author}{Vettorel, T.},
  \bibinfo{year}{2004}.
\newblock \bibinfo{title}{From weak to strong pinning {{I}}: A finite size
  study}.
\newblock \bibinfo{journal}{Eur. Phys. J. B} \bibinfo{volume}{38},
  \bibinfo{pages}{71--82}.
\newblock \DOIprefix\doi{10.1140/epjb/e2004-00101-6}.
\bibitem[{Thomson et~al.(1971)Thomson, Hsieh and Rana}]{thomson_lattice_1971}
\bibinfo{author}{Thomson, R.}, \bibinfo{author}{Hsieh, C.},
  \bibinfo{author}{Rana, V.}, \bibinfo{year}{1971}.
\newblock \bibinfo{title}{Lattice trapping of fracture cracks}.
\newblock \bibinfo{journal}{J. Appl. Phys.} \bibinfo{volume}{42},
  \bibinfo{pages}{3154--3160}.
\newblock \DOIprefix\doi{10.1063/1.1660699}.
\bibitem[{Violano and Afferrante(2021)}]{violano_roughness-induced_2021}
\bibinfo{author}{Violano, G.}, \bibinfo{author}{Afferrante, L.},
  \bibinfo{year}{2021}.
\newblock \bibinfo{title}{Roughness-induced adhesive hysteresis in self-affine
  fractal surfaces}.
\newblock \bibinfo{journal}{Lubricants} \bibinfo{volume}{9},
  \bibinfo{pages}{7}.
\newblock \DOIprefix\doi{10.3390/lubricants9010007}.
\bibitem[{Vollebregt(2014)}]{vollebregt_bound-constrained_2014}
\bibinfo{author}{Vollebregt, E.A.H.}, \bibinfo{year}{2014}.
\newblock \bibinfo{title}{The bound-constrained conjugate gradient method for
  non-negative matrices}.
\newblock \bibinfo{journal}{J. Optim. Theory Appl.} \bibinfo{volume}{162},
  \bibinfo{pages}{931--953}.
\newblock \DOIprefix\doi{10.1007/s10957-013-0499-x}.
\bibitem[{Wang and M{\"u}ser(2017)}]{wang_gauging_2017}
\bibinfo{author}{Wang, A.}, \bibinfo{author}{M{\"u}ser, M.H.},
  \bibinfo{year}{2017}.
\newblock \bibinfo{title}{Gauging {Persson} theory on adhesion}.
\newblock \bibinfo{journal}{Tribol. Lett.} \bibinfo{volume}{65},
  \bibinfo{pages}{103}.
\newblock \DOIprefix\doi{10.1007/s11249-017-0886-9}.
\bibitem[{Wang et~al.(2021)Wang, Zhou and M{\"u}ser}]{wang_modeling_2021}
\bibinfo{author}{Wang, A.}, \bibinfo{author}{Zhou, Y.},
  \bibinfo{author}{M{\"u}ser, M.H.}, \bibinfo{year}{2021}.
\newblock \bibinfo{title}{Modeling adhesive hysteresis}.
\newblock \bibinfo{journal}{Lubricants} \bibinfo{volume}{9},
  \bibinfo{pages}{17}.
\newblock \DOIprefix\doi{10.3390/lubricants9020017}.
\bibitem[{Wei et~al.(2010)Wei, He and Zhao}]{wei_effects_2010}
\bibinfo{author}{Wei, Z.}, \bibinfo{author}{He, M.F.}, \bibinfo{author}{Zhao,
  Y.P.}, \bibinfo{year}{2010}.
\newblock \bibinfo{title}{The effects of roughness on adhesion hysteresis}.
\newblock \bibinfo{journal}{J. Adhes. Sci. Technol.} \bibinfo{volume}{24},
  \bibinfo{pages}{1045--1054}.
\newblock \DOIprefix\doi{10.1163/016942409X12584625925222}.
\bibitem[{Wu(2010)}]{wu_jump--contact_2010}
\bibinfo{author}{Wu, J.J.}, \bibinfo{year}{2010}.
\newblock \bibinfo{title}{The jump-to-contact distance in atomic force
  microscopy measurement}.
\newblock \bibinfo{journal}{J. Adhes.} \bibinfo{volume}{86},
  \bibinfo{pages}{1071--1085}.
\newblock \DOIprefix\doi{10.1080/00218464.2010.519256}.
\bibitem[{Xia et~al.(2012)Xia, Ponson, Ravichandran and
  Bhattacharya}]{xia_toughening_2012}
\bibinfo{author}{Xia, S.}, \bibinfo{author}{Ponson, L.},
  \bibinfo{author}{Ravichandran, G.}, \bibinfo{author}{Bhattacharya, K.},
  \bibinfo{year}{2012}.
\newblock \bibinfo{title}{Toughening and asymmetry in peeling of heterogeneous
  adhesives}.
\newblock \bibinfo{journal}{Phys. Rev. Lett.} \bibinfo{volume}{108},
  \bibinfo{pages}{196101}.
\newblock \DOIprefix\doi{10.1103/PhysRevLett.108.196101}.
\bibitem[{Xia et~al.(2015)Xia, Ponson, Ravichandran and
  Bhattacharya}]{xia_adhesion_2015}
\bibinfo{author}{Xia, S.}, \bibinfo{author}{Ponson, L.},
  \bibinfo{author}{Ravichandran, G.}, \bibinfo{author}{Bhattacharya, K.},
  \bibinfo{year}{2015}.
\newblock \bibinfo{title}{Adhesion of heterogeneous thin films {{II}}: Adhesive
  heterogeneity}.
\newblock \bibinfo{journal}{J. Mech. Phys. Solids} \bibinfo{volume}{83},
  \bibinfo{pages}{88--103}.
\newblock \DOIprefix\doi{10.1016/j.jmps.2015.06.010}.
\bibitem[{Zappone et~al.(2007)Zappone, Rosenberg and
  Israelachvili}]{zappone_role_2007}
\bibinfo{author}{Zappone, B.}, \bibinfo{author}{Rosenberg, K.J.},
  \bibinfo{author}{Israelachvili, J.}, \bibinfo{year}{2007}.
\newblock \bibinfo{title}{Role of nanometer roughness on the adhesion and
  friction of a rough polymer surface and a molecularly smooth mica surface}.
\newblock \bibinfo{journal}{Tribol. Lett.} \bibinfo{volume}{26},
  \bibinfo{pages}{191}.
\newblock \DOIprefix\doi{10.1007/s11249-006-9172-y}.
\bibitem[{Zhou(2000)}]{zhou_critical_2000}
\bibinfo{author}{Zhou, Y.}, \bibinfo{year}{2000}.
\newblock \bibinfo{title}{Critical Dynamics of Contact Lines}.
\newblock Ph.D. thesis. Johns Hopkins University. \bibinfo{address}{{Baltimore,
  Maryland, USA}}.

\end{thebibliography}

\end{document}